\documentclass[twocolumn,showpacs,preprintnumbers,amsmath,amssymb,superscriptaddress,aps,prd,longbibliography]{revtex4-1}

\usepackage[pdftex]{graphicx} 
\usepackage{natbib}
\usepackage{booktabs}
\usepackage{color}
\usepackage{url}
\usepackage{float}
\usepackage{amsmath}
\usepackage{soul,xcolor}

\setstcolor{red}

\begin{document}
	
\title{Decoupling of static and dynamic criticality in a driven Mott insulator}

\author{A. de la Torre}
\affiliation{Department of Physics, California Institute of Technology, Pasadena, CA 91125, USA}
\affiliation{Institute for Quantum Information and Matter, California Institute of Technology, Pasadena, CA 91125, USA}
\author{K. L. Seyler}
\affiliation{Department of Physics, California Institute of Technology, Pasadena, CA 91125, USA}
\affiliation{Institute for Quantum Information and Matter, California Institute of Technology, Pasadena, CA 91125, USA}
\author{M. Buchhold}
\affiliation{Department of Physics, California Institute of Technology, Pasadena, CA 91125, USA}
\affiliation{Institute for Quantum Information and Matter, California Institute of Technology, Pasadena, CA 91125, USA}
\author{Y. Baum}
\affiliation{Department of Physics, California Institute of Technology, Pasadena, CA 91125, USA}
\affiliation{Institute for Quantum Information and Matter, California Institute of Technology, Pasadena, CA 91125, USA}
\author{G. Zhang}
\affiliation{Department of Physics, University of California, San Diego, La Jolla, California 92093, USA}
\author{N.J. Laurita}
\affiliation{Department of Physics, California Institute of Technology, Pasadena, CA 91125, USA}
\author{J.W. Harter}
\affiliation{Department of Physics, California Institute of Technology, Pasadena, CA 91125, USA}
\affiliation{Institute for Quantum Information and Matter, California Institute of Technology, Pasadena, CA 91125, USA}
\author{L. Zhao}
\affiliation{Department of Physics, California Institute of Technology, Pasadena, CA 91125, USA}
\affiliation{Institute for Quantum Information and Matter, California Institute of Technology, Pasadena, CA 91125, USA}
\author{I. Phinney}
\affiliation{Department of Physics, California Institute of Technology, Pasadena, CA 91125, USA}
\affiliation{Institute for Quantum Information and Matter, California Institute of Technology, Pasadena, CA 91125, USA}
\author{X. Chen}
\affiliation{Department of Physics, Boston College, Chestnut Hill, MA, 02467, USA}
\affiliation{Materials Department, University of California, Santa Barbara, CA, 93106, USA}
\author{S. D. Wilson}
\affiliation{Materials Department, University of California, Santa Barbara, CA, 93106, USA}
\author{G. Cao}
\affiliation{Department of Physics, University of Colorado, Boulder, CO 80309, USA}
\author{R. D. Averitt}
\affiliation{Department of Physics, University of California, San Diego, La Jolla, California 92093, USA}
\author{G. Refael}
\affiliation{Department of Physics, California Institute of Technology, Pasadena, CA 91125, USA}
\affiliation{Institute for Quantum Information and Matter, California Institute of Technology, Pasadena, CA 91125, USA}
\author{D. Hsieh}
\email[Corresponding author: ]{dhsieh@caltech.edu}
\affiliation{Department of Physics, California Institute of Technology, Pasadena, CA 91125, USA}
\affiliation{Institute for Quantum Information and Matter, California Institute of Technology, Pasadena, CA 91125, USA}

\date{\today}

\maketitle

\textbf{Dynamically driven interacting quantum many-body systems have the potential to exhibit properties that defy the laws of equilibrium statistical mechanics. A widely studied model is the impulsively driven antiferromagnetic (AFM) Mott insulator, which is predicted to realize exotic transient phenomena including dynamical phase transitions into thermally forbidden states \cite{Werner2012,Sandri2013,Tsuji2013} and highly non-thermal magnon distributions \cite{Walldorf2019}. However such far-from-equilibrium regimes, where conventional time-dependent Ginzburg-Landau descriptions fail, are experimentally challenging to prepare and to probe especially in solid state systems. Here we use a combination of time-resolved second harmonic optical polarimetry and coherent magnon spectroscopy to interrogate $n$-type photo-doping induced ultrafast magnetic order parameter (MOP) dynamics in the Mott insulator Sr$_2$IrO$_4$. We find signatures of an unusual far-from-equilibrium critical regime in which the divergences of the magnetic correlation length and relaxation time are decoupled. This violation of conventional thermal critical behavior arises from the interplay of photo-doping and non-thermal magnon population induced demagnetization effects. Our findings, embodied in a non-equilibrium “phase diagram”, provide a blueprint for engineering the out-of-equilibrium properties of quantum matter, with potential applications to terahertz spintronics technologies.} 

The low energy electronic structure of the strongly spin-orbit coupled Mott insulator Sr$_2$IrO$_4$ consists of a completely filled band of spin-orbital entangled pseudospin $J_{\rm eff} = 3/2$ states and a narrow half-filled band of $J_{\rm eff} = 1/2$ states, which splits into lower and upper Hubbard bands due to on-site Coulomb repulsion \cite{Kim2008}. Short-range AFM correlations between $J_{\rm eff} = 1/2$ moments within a layer are established well above room temperature \cite{Fujiyama2012,Vale2015}. But owing to weak interlayer exchange coupling ($J_c < 10^{-3} J_{ab}$) \cite{Vale2015,Porras2019}, three-dimensional (3D) long-range AFM ordering only occurs below a N\'eel temperature $T_{\rm N} = 230$ K. The ordered structure exhibits a weak in-plane ferromagnetic moment in each layer due to pseudospin canting, which is staggered along the $c$-axis (Fig. 1a). Resonant x-ray and neutron scattering measurements of La-doped Sr$_2$IrO$_4$ have shown that a small concentration of $n$-type carriers destroys $c$-axis ordering, leaving behind a paramagnetic state with short-range 2D AFM correlations \cite{Chen2015,Gretarsson2016,Pincini2017}. As such, $n$-type photo-doping, which may be simulated by optically pumping electrons from the electronically inert $J_{\rm eff} = 3/2$ band into the upper Hubbard band, provides a potential pathway to impulsively quench long-range AFM order.

Recently, both ultrafast optical and free-electron laser-based resonant x-ray scattering (RXS) methods have been used to probe photo-doping induced pseudospin dynamics in iridates \cite{Dean2016,Krupin2016,Afanasiev2019,Mazzonee2021,pastor2021}. In the case of Sr$_2$IrO$_4$, one study showed that it could be made magneto-optically active by aligning the canted moments in an external magnetic field \cite{Afanasiev2019}. Transient magnetization dynamics were then probed in the near-equilibrium weak photo-doping limit using time-resolved magneto-optical Kerr effect (MOKE) spectroscopy. The recovery time of the MOP was observed to diverge at $T_{\rm N}$, which was explained within a time-dependent Ginzburg-Landau framework. In another pair of studies, time-resolved RXS was used to probe MOP dynamics without a field in a far-from-equilibrium strong photo-doping regime \cite{Dean2016,Krupin2016}. A non-thermal melting of the MOP was achieved, but no critical dynamics were resolved across this phase transition.  

The behavior of a system near a dynamical phase boundary should be sensitive to small changes in excitation density. In order to resolve out-of-equilibrium critical dynamics, it is therefore imperative that the excitation density be uniform throughout the probed volume and finely sampled near the phase boundary. Since the pump excitation density invariably decays with depth into the sample, one should in principle probe exclusively within a small interval of depth to minimize the effects of averaging over different excitation densities, which might otherwise obscure features of criticality (see Supplementary sections S1 and S2). 
  
This condition is well met using an optical second harmonic generation rotational anisotropy (SHG-RA) technique (Fig. 1b), which was recently shown to be directly sensitive to the MOP in the topmost layer of Sr$_2$IrO$_4$ in zero magnetic field \cite{DiMatteo2016,Seyler2020}. We note that while an additional hidden order has been reported in Sr$_2$IrO$_4$ \cite{Jeong2017,Murayama2021}, there is no indication that SHG couples to any hidden order \cite{Seyler2020}, contrary to a previous interpretation \cite{Zhao2016}. In an SHG-RA experiment, light is focused obliquely onto the (001) face of Sr$_2$IrO$_4$ and the intensity of light reflected at twice the incident frequency $I(2\omega)$ is measured as the scattering plane is rotated about the surface normal. In equilibrium, SHG-RA patterns acquired above $T_{\rm N}$ exhibit four-fold ($C_4$) rotational symmetry, arising from a bulk electric-quadrupole (EQ) radiation from its centrosymmetric crystallographic point group \cite{Seyler2020}. Below $T_{\rm N}$, an additional surface magnetization-induced electric-dipole (ED) contribution turns on and interferes with the EQ contribution, lowering the symmetry of the SHG-RA pattern from $C_4$ to $C_1$ (Fig. 1c). 

To study the effects of $n$-type photo-doping, we developed time-resolved pump-probe SHG-RA using the apparatus depicted in Figure 1b. The pump beam, which was tuned on resonance with the $J_{\rm eff} = 3/2$ band to upper Hubbard band transition, and probe beam were focused within a single magnetic domain for our experiments. Figure 1d shows transient SHG-RA data acquired at the instant of pump excitation ($t$ = 0) as a function of pump fluence ($F$). In the un-pumped case ($F$ = 0), the $C_1$ symmetry of the SHG-RA pattern is manifested through the presence of a dominant intensity lobe, indicating a finite ED contribution and thus a finite MOP. As $F$ increases, the intensity of the dominant lobe decreases linearly until it plateaus above a critical fluence $F_{\rm c} \approx 0.9$ mJ/cm$^2$. Beyond $F_{\rm c}$, the non-magnetic $C_4$ symmetry is fully restored to the SHG-RA pattern, signaling a collapsed MOP. The slightly higher $C_4$-EQ yield from the pump-induced compared to heating-induced non-magnetic state likely arises from un-relaxed bulk magneto-elastic deformations that will be discussed in more detail below. The prompt change in the SHG-RA pattern within the time resolution of our instrument ($< 200$ fs) for all fluences is consistent with a photo-doping scenario. Specifically, AFM order is suppressed by magnetic defects that are left in the wake of propagating doublons (Fig. 1b), which are generated on the time scale of the nearest-neighbor hopping ($< 20$ fs) \cite{Takahashi2002,Piovera2016,Afanasiev2019}. Moreover, the observed critical fluence corresponds to an excitation density ($n_{\rm ex}$) of approximately 0.05 per iridium site, which is close to the reported critical La-doping level to suppress AFM order \cite{Chen2015,Gretarsson2016}, suggestive of an optical $n$-type doping induced quench mechanism. 

More detailed comparisons between the properties of Sr$_2$IrO$_4$ in and out-of-equilibrium can be made by mapping the fluence versus temperature magnetic “phase diagram”. This is accomplished by collecting an array of SHG-RA patterns immediately after optical excitation across different starting temperatures and pump fluences and then tracking the onset of the ED contribution. For $F$ = 0, the ED contribution emerges below $T_{\rm N}$ as expected (Fig. 2a). With increasing $F$, this transition shifts monotonically to lower temperatures, indicating that $F_{\rm c}$ increases upon cooling. By plotting $T_{\rm N}$ as a function of $n_{\rm ex}$, or equivalently $F_{\rm c}$ as a function of $T$, a sharp out-of-equilibrium magnetic phase boundary is identified. Despite an instantaneous electronic temperature that already far exceeds $T_{\rm N}$ at $F \approx 0.2$ mJ/cm$^2$ based on the electronic specific heat of Sr$_2$IrO$_4$ (see Supplementary section S3), the photo-doping and La-doping phase boundaries nearly coincide over the small La-doping range where AFM order exists in equilibrium \cite{Chen2015,Gretarsson2016} (Fig. 2b). This indicates that heat transfer from the charge to pseudospin subsystem is negligible near $t$ = 0. Since chemical doping differs from photo-doping in non-trivial ways, including the introduction of disorder that is known to nucleate non-magnetic metallic puddles in La-doped Sr$_2$IrO$_4$ \cite{Chen2015,Battisti2017}, we refrain from drawing more detailed comparisons.   

At later time delays, the charge, pseudospin and lattice subsystems are typically expected to thermalize and thus the out-of-equilibrium phase diagrams mapped at $t \approx 0$ and $t \gg 0$ should be different. Figure 2c shows the intensity of the dominant SHG-RA lobe over a range of fluence values, converted into effective temperatures using the total specific heat of Sr$_2$IrO$_4$ (see Methods), acquired at $t = 10$ ps, which far exceeds the reported charge and lattice relaxation timescales but is much shorter than the timescale for heat to escape the probed region \cite{Hsieh2012,Piovera2016}. At low fluences we find excellent agreement with the un-pumped temperature dependence data, confirming a pure optical heating effect. Surprisingly however, there is increasing bifurcation of the curves above $F \approx 0.8$ mJ/cm$^2$, leading to a growing mismatch between the out-of-equilibrium phase boundary mapped at $t = 10$ ps and that calculated assuming pure optical heating as a function of fluence (Fig. 2d). This indicates that thermalization of the pseudospin subsystem is impeded at high fluences.    

A hallmark of the La-doping induced paramagnetic state is the persistence of short-range intralayer AFM correlations, manifested through remnant dispersive 2D magnons detected by resonant inelastic x-ray scattering \cite{Gretarsson2016,Pincini2017}. To search for similar intralayer correlations in the photo-induced paramagnetic phase of Sr$_2$IrO$_4$, we leverage the fact that the zone center 2D magnon mode with $B_{2g}$ symmetry is weakly gapped ($ \approx 2$ meV) and Raman active \cite{Gim2016, Gretarsson2017}. This allows the magnon to be coherently excited by our pump pulse via impulsive stimulated Raman scattering and optically tracked in the time domain using ultrafast MOKE spectroscopy \cite{Seifert2019}. Figures 2e,f show typical SHG and MOKE transients acquired in the high pump fluence ($F \gg F_{\rm c}$) regime. Despite the complete collapse of the MOP over the displayed 10 ps time window (Fig. 2e), coherent oscillations of the infinite wavelength $B_{2g}$ magnon continue to be supported (Fig. 2f). This provides clear evidence of a significant 2D magnetic correlation length in the transient paramagnetic state, suggesting a non-thermal analogue of La-doped Sr$_2$IrO$_4$ is realized. These results are consistent with and complementary to a time-resolved RXS study of Sr$_2$IrO$_4$ \cite{Dean2016}, which showed that photo-doping strongly suppresses an AFM Bragg peak but does not alter the main high energy and short wavelength features of the intralayer magnon spectrum (see Supplementary section S4). 

Having comprehensively mapped the transient magnetic phase diagram of Sr$_2$IrO$_4$, we now finely examine the out-of-equilibrium critical dynamics. Figures 3a,b show the time-dependent change in intensity $\Delta I(2\omega)$ of the dominant SHG-RA lobe acquired at $T = 80$ K over a range of pump fluences. For $F \ll F_{\rm c}$ (Fig. 3a), we observe a rapid drop in $\Delta I(2\omega)$ at $t$ = 0 signifying a reduction of the MOP. This is followed by an exponential recovery with rise time $\tau \approx 1 – 2$ ps towards a value negatively offset from the $t < 0$ intensity, which we showed (Fig. 2c,d) represents a thermalized state with a slightly elevated temperature. As $F$ increases, $\tau$ rises slightly but exhibits no discontinuity at $F_{\rm c}$ despite the MOP vanishing. However as $F$ exceeds $F_{\rm c}$, the point where $C_4$ symmetry is restored, $\tau$ continues to grow and eventually diverges around $F^* \approx 1.6$ mJ/cm$^2$ (Fig. 3b,c). Similar dynamics are observed if instead $F$ is kept fixed and $T$ is varied (inset Fig. 3a). These data reveal that the divergence of the magnetic correlation length and relaxation time occur along separated critical lines in the out-of-equilibrium phase diagram marked by $F_{\rm c}$ and $F^*$ respectively (inset Fig. 3c). Such a decoupling is forbidden in equilibrium and is also not observed across dynamical phase transitions in weakly correlated ferromagnets \cite{Tengdin2018} and charge density wave systems \cite{ZongPRL}.    

We note the presence of a subtle exponential drop just after $t$ = 0 that becomes more pronounced at large $F$. This causes $\Delta I(2\omega)$ to drop below the instantaneous EQ value (gray bar in {Fig. 1} and Figs. 3a,b and ultimately plateau at the high temperature EQ value upon reaching $F^*$. Given that magneto-elastic deformations are known to occur below $T_{\rm N}$ in Sr$_2$IrO$_4$ \cite{Porras2019} and can cause small changes in the EQ response, a possible origin of this exponential component is lattice relaxation following impulsive suppression of the MOP. Therefore, only at large $F$ when $\tau$ sufficiently exceeds the relevant phonon timescales do the magneto-elastic deformations have time to fully relax (see Supplementary section S5). Although a direct confirmation awaits time-resolved crystallography measurements, the main conclusions of our work are independent of this interpretation.

To check whether the divergence of $\tau$ simply results from a prolonged photo-dopant lifetime, we performed simultaneous transient linear reflectivity ($\Delta R/R$) and SHG-RA measurements below $T_N$ to directly compare the charge and pseudospin dynamics. As shown in Figures 4a-c, the generation of photo-dopants and their subsequent suppression of the MOP both occur within the time resolution of our experiment. However, unlike the SHG response, we observe no saturation in the amplitude of $\Delta R/R$ and no significant change in its characteristic recovery time ($\tau_0 \approx 1$ ps) as a function of fluence (inset Fig. 4a), demonstrating that the slow dynamics observed at $F^*$ occur exclusively in the pseudospin sector. This suggests that $\tau$ must be governed by the relaxation of excess magnons that are emitted upon doublon decay. One possibility is that the magnons rapidly thermalize to a higher temperature and that $\tau$ represents the timescale for the hot pseudospin subsystem to cool via heat exchange with the lattice and charge subsystems. However, a three-temperature model describing this process predicts that magnons should cool faster at higher excitation densities. Moreover, it fails to produce any diverging timescale (see Supplementary sections S3 and S6). This points to the alternative possibility that $\tau$ represents the timescale for the pseudospin subsystem to internally thermalize.

The rate-limiting step for recovering long-range magnetic order is the establishment of $c$-axis correlations (Fig. 2e,f). Therefore $\tau$ must be set by the thermalization time of interlayer $c$-axis magnons. However the large mismatch between $J_c$ ($\approx$ 10 $\mu$eV)\cite{Vale2015,Porras2019} and both the charge gap ($ \approx 250$ meV)\cite{delatorre2015} and lowest optical phonon energy ($\approx 10$ meV)\cite{Gretarsson2017} of Sr$_2$IrO$_4$ should impede $c$-axis magnon thermalization via charge or phonon excitation processes, leaving magnon-magnon scattering as the dominant thermalization channel. Building on the seminal work of Hohenberg and Halperin \cite{HohenbergHalperin1977}, we study the critical dynamics after doublon decay of this effectively closed $c$-axis magnon subsystem via a general stochastic Langevin equation for a real order parameter in one spatial dimension (see Supplementary section S7):   

\begin{widetext}
\begin{equation}
\partial_t\varphi_k\left(t\right)= -\left({k^2+\tau}_0^{-1}\right)\varphi_k\left(t\right)-\lambda\int\frac{dk_1}{2\pi}\int{\frac{dk_2}{2\pi}{\ \varphi}_{k_1}\left(t\right)\varphi_{k_2}\left(t\right)\varphi_{{k-k}_1-k_2}\left(t\right)+\xi_k\left(t\right)}
\end{equation}
\end{widetext}

\noindent where $\varphi_k\left(t\right)$ is the $k$-th Fourier component of the MOP field $\varphi\left(z,t\right)$ that parameterizes the ordering of the intralayer N\'eel vector along the $c$-axis (Fig. 4d), $\tau_0^{-1}$ is a temperature-dependent mass term that is equivalent to the thermalization rate in a linearized Boltzmann equation \cite{Sieberer2016}, and $\lambda$ parameterizes the interaction between magnons. Going beyond conventional time-dependent Ginzburg-Landau descriptions, we introduce a Gaussian noise term $\xi_k\left(t\right)$ with $\langle \xi_k(t) \xi_{k'}(t')\rangle =\delta (t-t') \delta (k+k')(k^2+\tau_0^{-1})[2n_k(t)+1]$ that imprints a non-equilibrium fluctuation-dissipation relation for some general magnon distribution function $n_k\left(t\right)$, in which the modes transverse to the $c$-axis are assumed to relax much faster and are thus integrated out \cite{Dolgirev2020}. To leading order, the equation of motion for the MOP $\Phi\left(t\right)=\langle \varphi_{k = 0}(t)\rangle$ is given by $\partial_t\Phi\left(t\right)=-\tau^{-1}\Phi\left(t\right)-\lambda{\Phi\left(t\right)}^3$, and the Boltzmann equation for the magnon distribution is given by $\partial_tn_k\left(t\right)=-\left|\tau^{-1}\right|\left[n_k\left(t\right)-n_k^T\right]$, where $\tau^{-1}=\tau_0^{-1}-\lambda N_{\rm mag}$ and $n_k^T$ is the Bose distribution function corresponding to the final equilibrated magnon temperature $T$. These equations show that, in a closed system, the rate at which the MOP recovers to its equilibrium value at $T$ via magnon thermalization is a decreasing function of the total number of excited magnons ($N_{\rm mag}$). This provides an explanation for all key features of the data in Fig. 3c. In the low pump fluence limit where $N_{\rm mag}\ll1/\tau_0\lambda$, $\tau$ approaches a lower bound that is naturally set by the photo-dopant decay time $\tau_0$. As $N_{\rm mag}$ grows with fluence, $\tau$ exhibits a power law increase following the reduced curvature of the Mexican hat potential. Finally once $N_{\rm mag}$ reaches $1/\tau_0\lambda$ at the critical fluence $F^*$, $\tau$ diverges even though the total energy of the magnetic subsystem remains sub-critical (see Supplementary section S7). For $F \geq F^*$, thermalization can still eventually occur through the weak coupling between $c$-axis magnons and phonons, which is likely responsible for the slow ($> 100$ ps) MOP recovery observed by time-resolved RXS measurements at high fluences \cite{Dean2016,Krupin2016}. Ultimately, on even longer timescales, the system cools back to the initial temperature via heat diffusion out of the pumped region (Fig. 4d). 

Altogether, our results provide experimental signatures that static and dynamic critical behavior can be decoupled in a Mott antiferromagnet upon driving due to a subtle interplay of photo-doping and magnon thermalization processes (Fig. 4d). The hierarchy of magnetic, phononic and charge gap energy scales that enables this behavior in Sr$_2$IrO$_4$ is typical for layered transition metal oxide based AFM Mott insulators including the high-$T_c$ superconducting cuprates, suggesting that our reported phenomena may be operational in numerous layered quantum materials. This raises the intriguing prospect of fine-tuning parameters such as $\tau_0$, $F_{\rm c}$ and $F^*$ to optimize desired out-of-equilibrium properties. For example, extending $\tau_0$ may be a pathway for realizing metastable electronic instabilities in a photo-doped Mott insulator, while the combination of a small $\tau_0$ and $F_{\rm c}$ with a large $F^*$ would be favorable for low-power high-speed AFM switching applications.

\section{Acknowledgements}

We thank L. Balents, A. Cavalleri, S. K. Cushing, E. Demler, S. Di. Matteo, M. Endres, B. Fine, N. Gedik, D. Kennes, O. Mehio, H. Ning, M. Norman and M. Sentef for useful discussions. This work is supported by ARO MURI Grant No. W911NF-16-1-0361. D.H. also acknowledges support for instrumentation from the David and Lucile Packard Foundation and from the Institute for Quantum Information and Matter (IQIM), an NSF Physics Frontiers Center (PHY-1733907). A.d.l.T. acknowledges support from the Swiss National Science Foundation through an Early Postdoc Mobility Fellowship (P2GEP2$_165044$). M.B. acknowledges support from the Alexander von Humboldt foundation. N.J.L. acknowledges support from an IQIM Fellowship. G.C. acknowledges NSF support via a grant DMR-1712101.  

\section{Competing Interests}

The authors declare no competing interests.

\section{Methods}

\subsection{Sample growth}
Single crystals of Sr$_2$IrO$_4$ were grown using a self-flux technique from off-stoichiometric quantities of IrO$_2$, SrCO$_3$ and SrCl$_2$. The ground mixtures of powders were melted at 1470$^{\circ}$C in partially capped platinum crucibles. The soaking phase of the synthesis lasted for $> 20$ hours and was followed by a slow cooling at 2$^{\circ}$C/hr to reach 1400$^{\circ}$C. From this point the crucible is brought to room temperature through a rapid cooling at a rate of 100$^{\circ}$C/hr. Crystals were cleaved along the (001) face just prior to measurements and immediately pumped down to a pressure $< 5 \times 10^{-6}$ torr in an optical cryostat.

\subsection{Time-resolved spectroscopic probes}
\textit{Time-resolved SHG-RA}: Experiments were performed by splitting light from a regeneratively amplified Ti:sapphire laser, which produces 100 fs pulses with central wavelength $\lambda$ = 800 nm at a repetition rate of 100 kHz, into pump and probe arms. The pump beam feeds an optical parametric amplifier, whose 1400 nm output beam is linearly polarized and sent through a delay line before being focused at normal incidence onto a 90 $\mu$m spot (FWHM) located within a single magnetic domain of a Sr$_2$IrO$_4$ crystal. As shown in Fig. 1b, the circularly polarized probe laser pulse ($\lambda$ = 800 nm) traverses a linear input polarizer, phase mask and collimating lens and is focused at oblique incidence ($\approx 10^{\circ}$) onto a 30 $\mu$m spot within the pumped sample area using an objective lens. The reflected SHG pulse ($\lambda$ = 400 nm) is re-collimated by the objective lens, linearly polarized by an output polarizer and then deflected onto a CCD camera by a triad of dichroic mirrors, which measures the SHG intensity at a fixed time delay with respect to the pump. The probe fluence was fixed to $1.5$ mJ/cm$^2$. A complete instantaneous SHG-RA pattern is acquired by mechanically co-rotating the input polarizer, phase mask and output polarizer, which implements a rotation of the probe scattering plane about the sample surface normal, and then projecting the SHG light reflected at each scattering plane angle onto different positions along a circular locus of points on the CCD camera  [see Ref. \cite{Harter2015} for further details]. The procedure for converting raw CCD data to the polar SHG-RA plots shown in the main text is described in the Supplementary section S8. The SHG dynamics shown in Fig. 3 were reproduced under different polarization geometries and different scattering plane angles as shown in Supplementary section S9.

\textit{Time-resolved linear reflectivity}: Transient reflectivity measurements were performed using exactly the same setup as that used for time-resolved SHG-RA as described above. Identical experimental conditions were preserved for the reflectivity and SHG-RA measurements, with the only change between experimental runs being the spectral filtering optics before the CCD camera to either isolate the 800 nm or 400 nm output. 

\textit{Time-resolved MOKE}: Experiments were performed by splitting light from a regeneratively amplified Ti:sapphire laser (repetition rate 1 kHz; pulse energy 6 mJ; pulse duration 100 fs; central wavelength $\lambda$ = 800 nm) into pump and probe arms. The pump arm feeds a collinear dual-stage optical parametric amplifier, whose 1300 nm output beam is circularly polarized and then focused onto the sample at normal incidence. The probe arm ($\lambda$ = 800 nm) is linearly S-polarized and focused near normal incidence onto the sample. To measure the polarization rotation (Kerr) angle of the reflected probe light, the reflected probe beam is sent through a half wave plate followed by a Wollaston prism. These optics are oriented such that in the absence of a pump beam ($\theta_k = 0$, $F = 4.6$ mJ/cm$^2$), two equally intense beams of orthogonal linear polarization are produced, which are fed into the two input channels of a balanced photo-detector. 

\subsection{Conversion from $F$ to $n_{\rm ex}$}
The number of photo-excitations $n_{\rm ex}$ created by the pump beam was calculated using the following equation:
\begin{equation}
n_{\rm ex}= \frac{F\left(1-R\right)V_{\rm u.c.}}{N_{\rm Ir}E_{\rm ph}\delta}
\end{equation}
where $F$ is the pump fluence, $R$ = 0.2 and $\delta$ = 100 nm are the sample reflectivity and penetration depth respectively at the pump wavelength $\lambda$ = 1400 nm \cite{Lee2012,Nichols2013}, $V_{\rm u.c.}$ is the unit cell volume of Sr$_2$IrO$_4$, $N_{\rm Ir}$ is the number of Ir atoms per unit cell, and $E_{\rm ph}$ the pump photon energy (0.89 eV).  

\subsection{Exponential fitting procedure for linear reflectivity}
Transient linear reflectivity curves were fit to a double exponential function of the form 
$\Delta R/R=Ae^{-t/\tau_0}+Be^{-t/\tau_1}+C$. A high quality of fit is achieved as shown in Fig. 4a. The timescales extracted for $\tau_0$ ($\approx$ 1 ps) and $\tau_1$ ( $\approx$ 200 fs) agree with the two exponential timescales associated with the decay of photo-induced midgap states observed by ultrafast  angle-resolved photoemission spectroscopy measurements \cite{Piovera2016}, attributed to acoustic phonon and optical phonon mediated photo-carrier relaxation processes respectively. Since the rate limiting step for photo-carrier relaxation is $\tau_0$, that is what we plot in the inset of Fig. 4a. We note that Ref. \cite{Piovera2016}, also reported an absence of any fluence dependence for $\tau_0$, which is consistent with our observations (inset Fig. 4a). 
%

\begin{figure*}[!ht]
    \includegraphics[width=\textwidth]{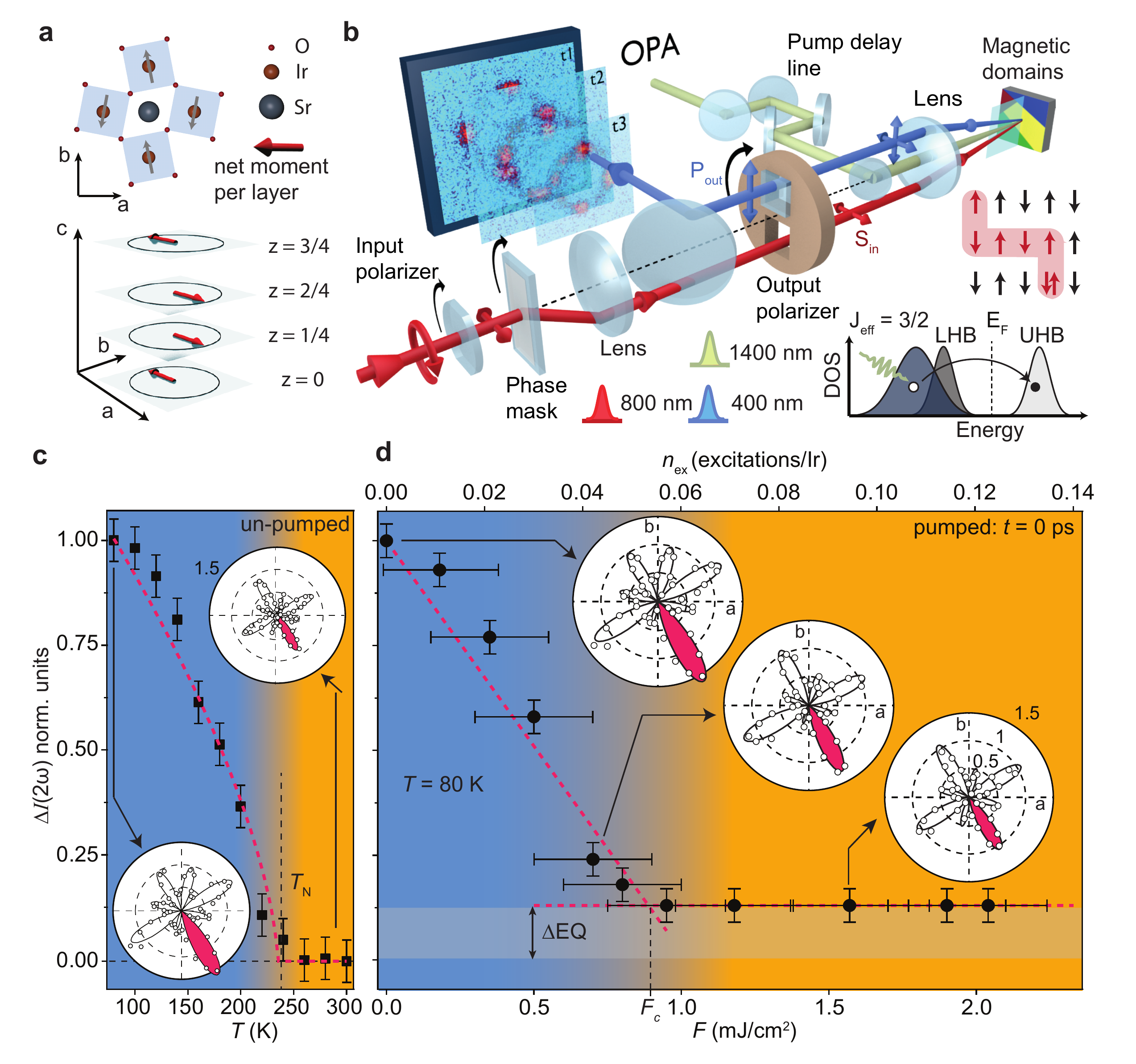}
	\caption{\textbf{Instantaneous photo-doping response of magnetic order in Sr$_2$IrO$_4$.} \textbf{a}, Intralayer magnetic order of Sr$_2$IrO$_4$. Each layer possesses a non-zero net moment due to pseudospin canting, which are staggered along the $c$-axis. \textbf{b}, Schematic of the time-resolved SHG-RA setup. A circularly polarized probe laser pulse ($\lambda$ = 800 nm) traverses a linear polarizer, phase mask, collimating lens and is focused onto the sample at oblique incidence using an objective lens. The reflected SHG pulse ($\lambda$ = 400 nm) is re-collimated by the objective, linearly polarized and deflected onto a CCD camera by a dichroic mirror. Different scattering plane angles are accessed by mechanically co-rotating select optics. Data shown throughout the manuscript are acquired in the P$_{\rm in}$-S$_{\rm out}$ polarization channel (see Methods). A time-delayed linearly polarized pump pulse resonant with the $J_{\rm eff} = 3/2$ to upper Hubbard band energy ($\lambda$ = 1400 nm) is focused normally onto the sample. Regions of different color on the sample represent distinct magnetic domains identified by SHG imaging. Inset: schematic of doublon generation and propagation. \textbf{c}, Temperature dependence of the change in SHG intensity $\Delta I(2\omega)$ relative to $T$ = 300 K for an un-pumped sample acquired at the angle of maximum intensity (pink). Dashed line is a guide to the eye. \textbf{d}, Pump fluence dependence of $\Delta I(2\omega)$ at $t$ = 0 and $T$ = 80 K with corresponding SHG-RA patterns. Fits to the data below (above) $F_{\rm c}$ including both EQ and ED terms (solely EQ term) are overlaid. Height of the gray bar represents the change in EQ SHG intensity with respect to the equilibrium value as discussed in the main text. Horizontal error bars represent the systematic error in calculating $n_{\rm ex}$.}
	\label{f1}
\end{figure*}

\begin{figure*}[!ht]
	\includegraphics[width=\textwidth]{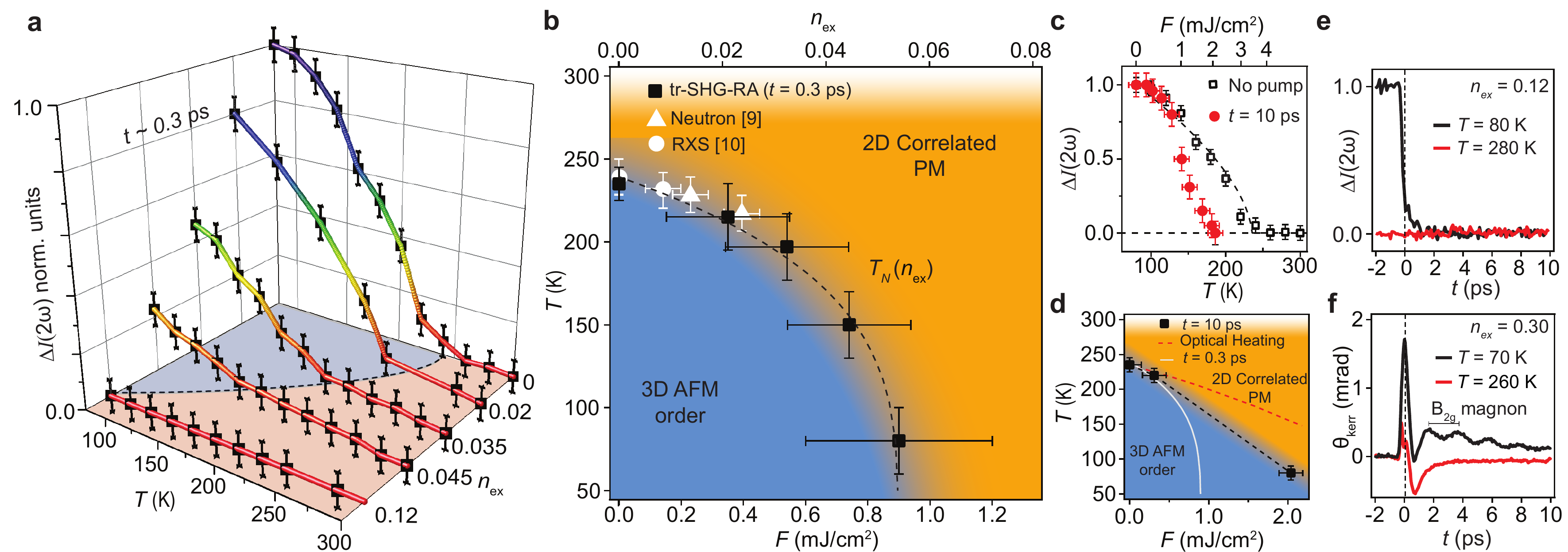}
	\caption{\textbf{Out-of-equilibrium magnetic phase diagrams.} \textbf{a}, Temperature dependence of $\Delta I(2\omega)$ at $t$ = 0.3 ps for select pump excitation densities $n_{\rm ex}$ and, \textbf{b}, the out-of-equilibrium phase diagram mapped using such datasets (at $t$ = 0.3 ps). The reported La-doping dependence of $T_{\rm N}$ is overlaid. Horizontal error bars represent the systematic error in calculating $n_{\rm ex}$. \textbf{c}, Comparison of the temperature dependence of $\Delta I(2\omega)$ for an un-pumped sample with the fluence dependence of $\Delta I(2\omega)$ collected at $t$ = 10 ps and $T$ = 80 K, where $F$ is converted into an effective temperature by assuming all of the pump energy goes into quasi-equilibrium heating. \textbf{d}, The out-of-equilibrium phase diagram mapped at $t$ = 10 ps (black dashed line). Overlaid are the phase boundary reproduced from panel \textbf{b} (white line) and that calculated assuming all of the pump energy goes into quasi-equilibrium heating (red line). \textbf{e}, High fluence $\Delta I(2\omega)$ and \textbf{f}, MOKE transients acquired below and above $T_{\rm N}$.}
	\label{f2}
\end{figure*}

\begin{figure*}[htb]
	\includegraphics[width=\textwidth]{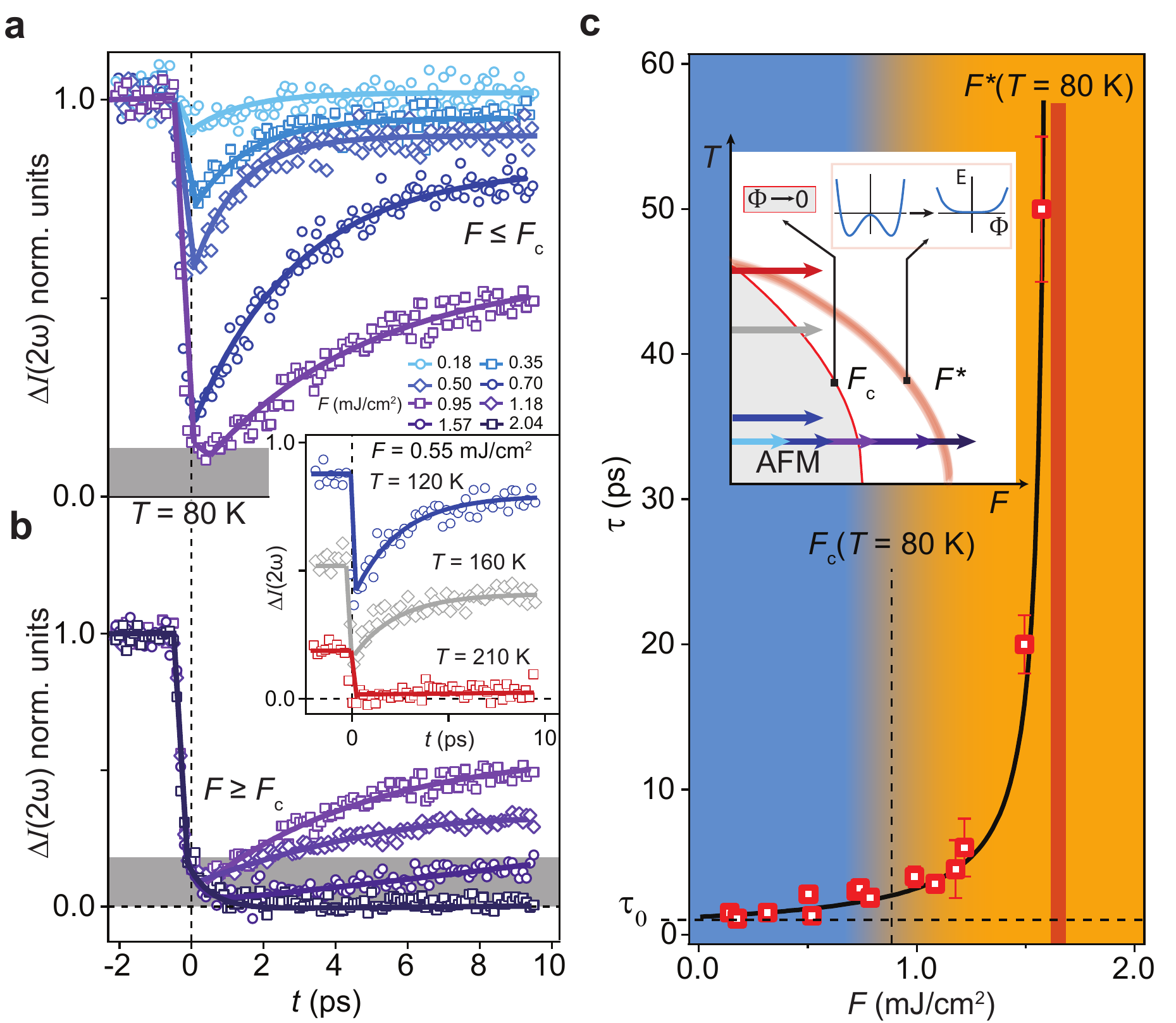}
	\caption{\textbf{Out-of-equilibrium critical dynamics of the magnetic order parameter}. $\Delta I(2\omega)$ transients acquired at $T$ = 80 K for select fluences \textbf{a}, below and \textbf{b}, above $F_{\rm c}$. The height of the gray bar denotes the excess EQ SHG intensity at $t = 0$ as discussed in the main text. Inset shows analogous data acquired at fixed $F = 0.55$ mJ/cm$^2$ for different temperatures. Solid lines are fits of the recovery to a single exponential function plus a constant determined by the long time offset of the SHG intensity. \textbf{c}, Plot of fitted exponential relaxation times versus fluence superposed with a fit to the function $\tau=\tau_0(1 – F/F^*)^{-1}$, with $\tau_0$ fixed to 1.2 ps by the transient reflectivity data (Fig. 4a). The values of $F_{\rm c}$ and $F^*$ at $T$ = 80 K based on the data in panels \textbf{a}, \textbf{b} and the fit (black line) respectively are explicitly marked. The decoupling of the divergence in magnetic correlation length (defined by $F_{\rm c}$) from the divergence in relaxation time (defined by $F^*$) is illustrated in the inset based on the data in panels \textbf{a} and \textbf{b}. The arrows are color coded in correspondence to the curves in panels \textbf{a} and \textbf{b} and point to the location in the out-of-equilibrium phase diagram being accessed. Schematics show the the out-of-equilibrium free energy landscape after doublon decay based on the Langevin theory described in the text.} 
	\label{f3}
\end{figure*}

\begin{figure*}[htb]
    \includegraphics[width=\textwidth]{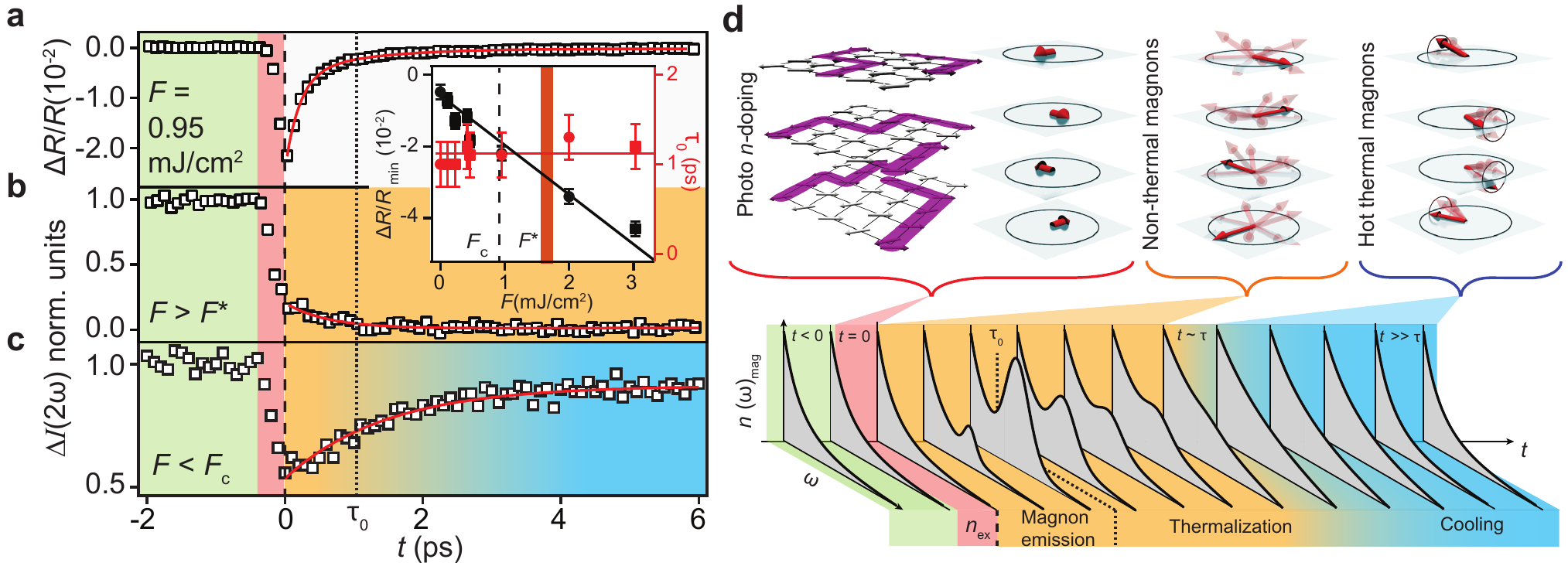}
	\caption{\textbf{Magnon population induced critical slowing down.} Simultaneous measurements of the \textbf{a}, transient reflectivity and \textbf{b-c}, transient SHG intensity acquired under identical experimental conditions (except for fluence) at $T$ = 80 K. Inset of panel \textbf{a}: fluence dependence of the intensity minimum and characteristic recovery time $\tau_0$ extracted through double exponential fits (red curve) to the $\Delta R/R$ data (see Methods). \textbf{d}, Schematic of the complete temporal evolution of (top) the real space $c$-axis magnetic correlations and (bottom) the magnon distribution function $n\left(\omega\right)$ versus energy $\omega$ following pump excitation. Over the time window $0<t\lesssim\tau$ (shaded orange), $n\left(\omega\right)$ departs from Bose-Einstein form.}
	\label{f4}
\end{figure*}
\newpage
\pagenumbering{gobble}
\begin{figure}
   \vspace*{-2cm}
   \hspace*{-2cm}
    \centering
    \includegraphics[page=1]{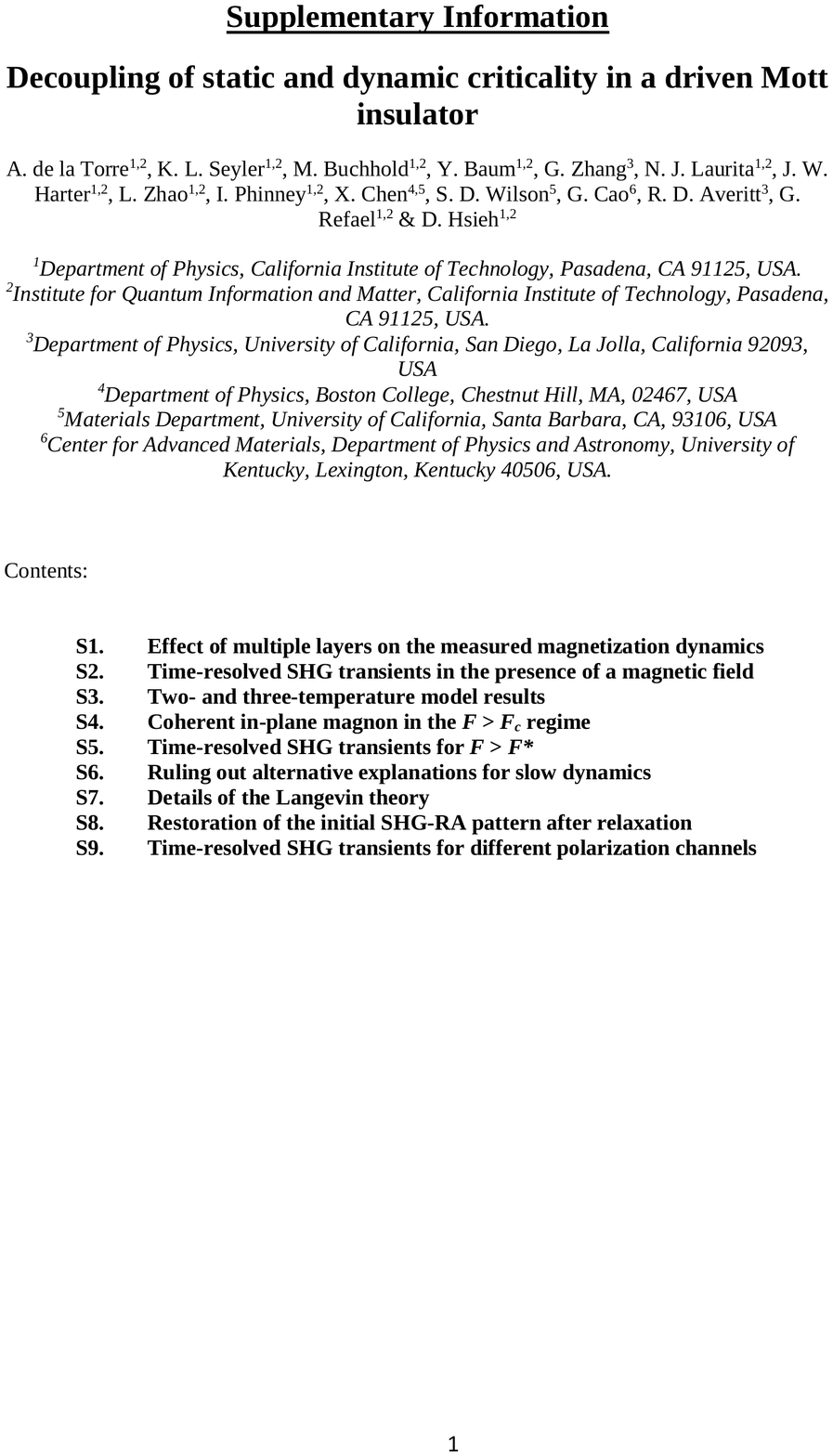}

\end{figure}
\begin{figure}
   \vspace*{-2cm}
   \hspace*{-2cm}
    \centering
    \includegraphics[page=2]{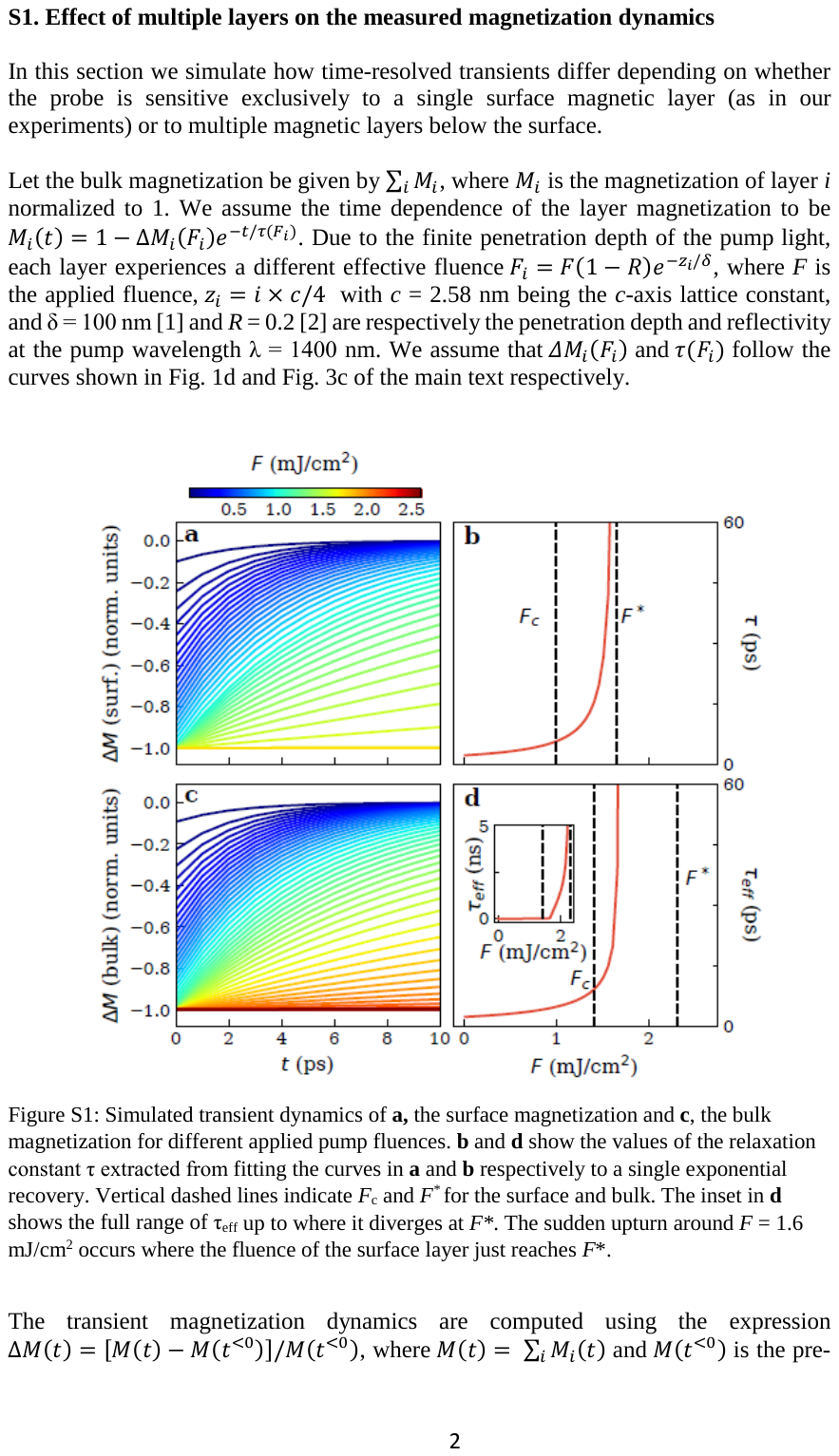}

\end{figure}
\begin{figure}
   \vspace*{-2cm}
   \hspace*{-2cm}
    \centering
    \includegraphics[page=3]{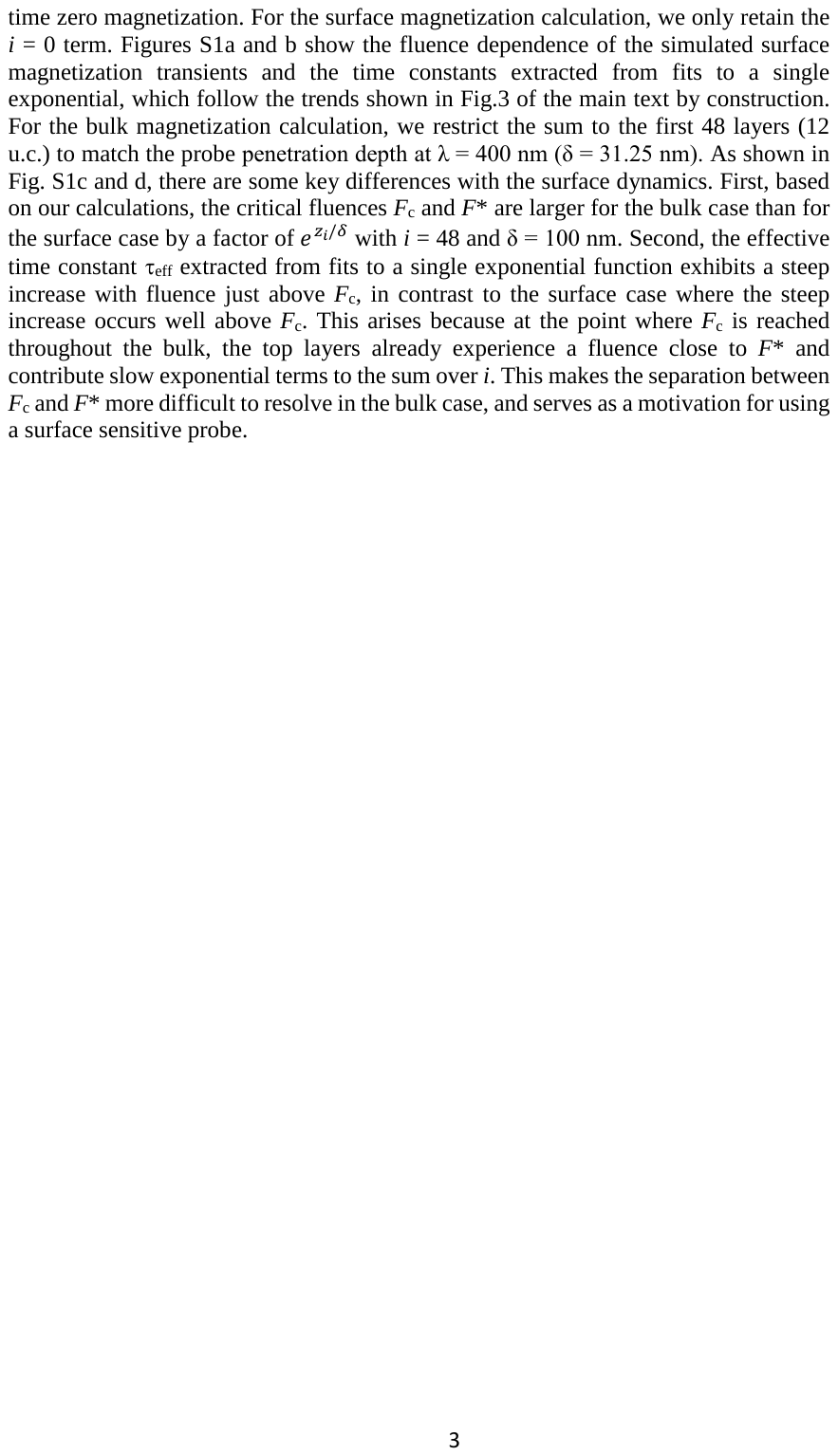}

\end{figure}
\begin{figure}
   \vspace*{-2cm}
   \hspace*{-2cm}
    \centering
    \includegraphics[page=4]{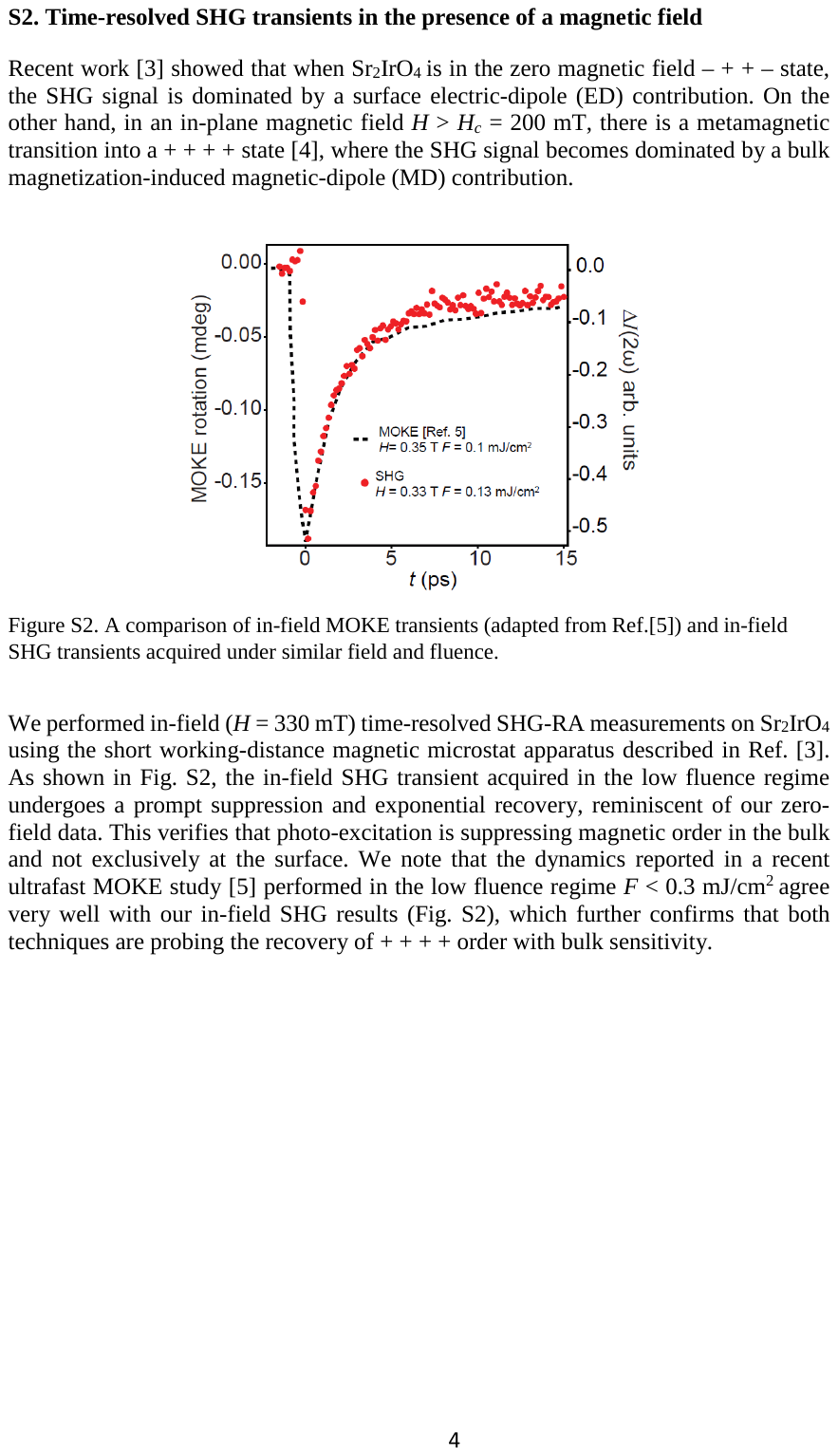}

\end{figure}
\begin{figure}
   \vspace*{-2cm}
   \hspace*{-2cm}
    \centering
    \includegraphics[page=5]{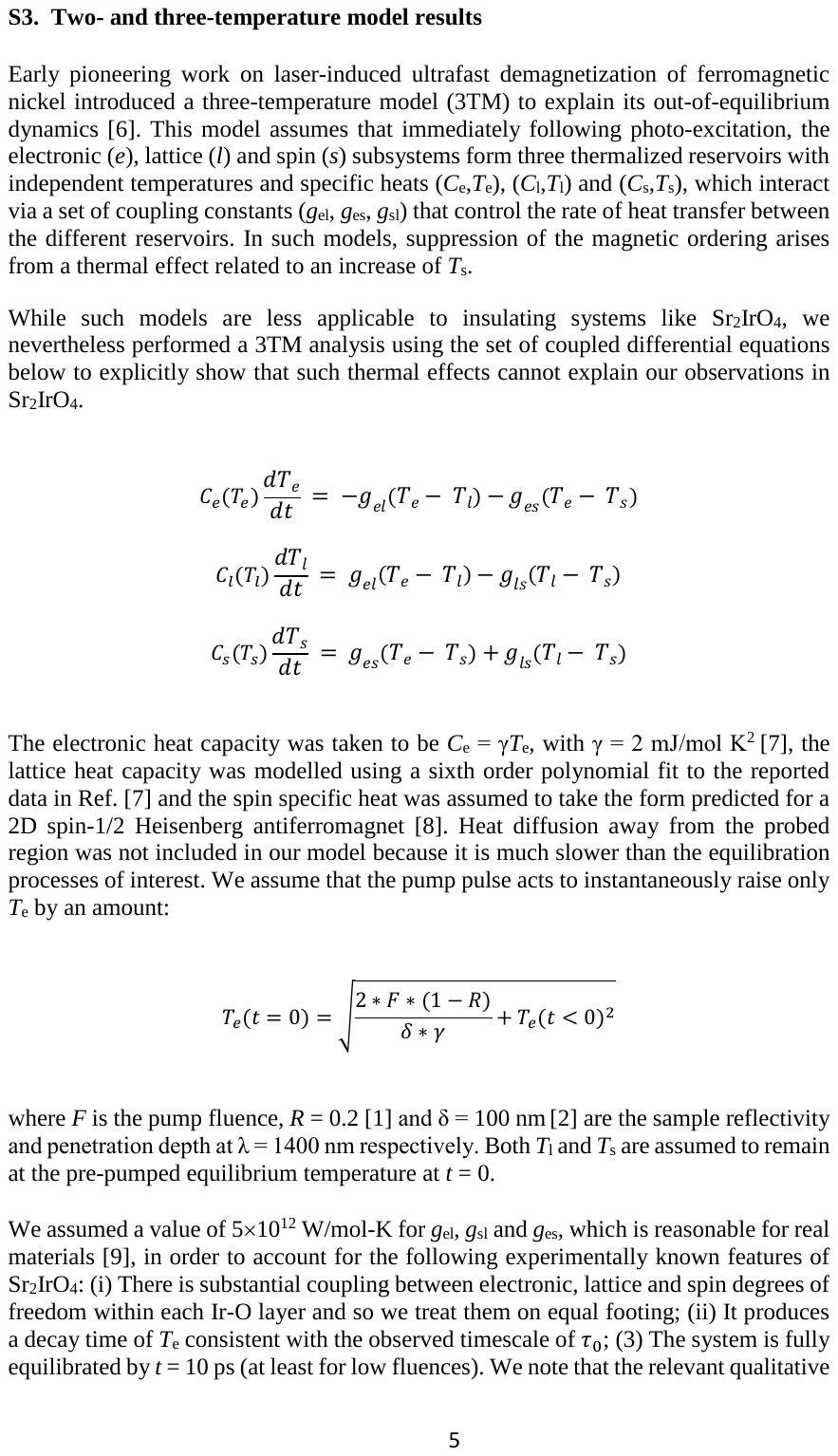}

\end{figure}
\begin{figure}
   \vspace*{-2cm}
   \hspace*{-2cm}
    \centering
    \includegraphics[page=6]{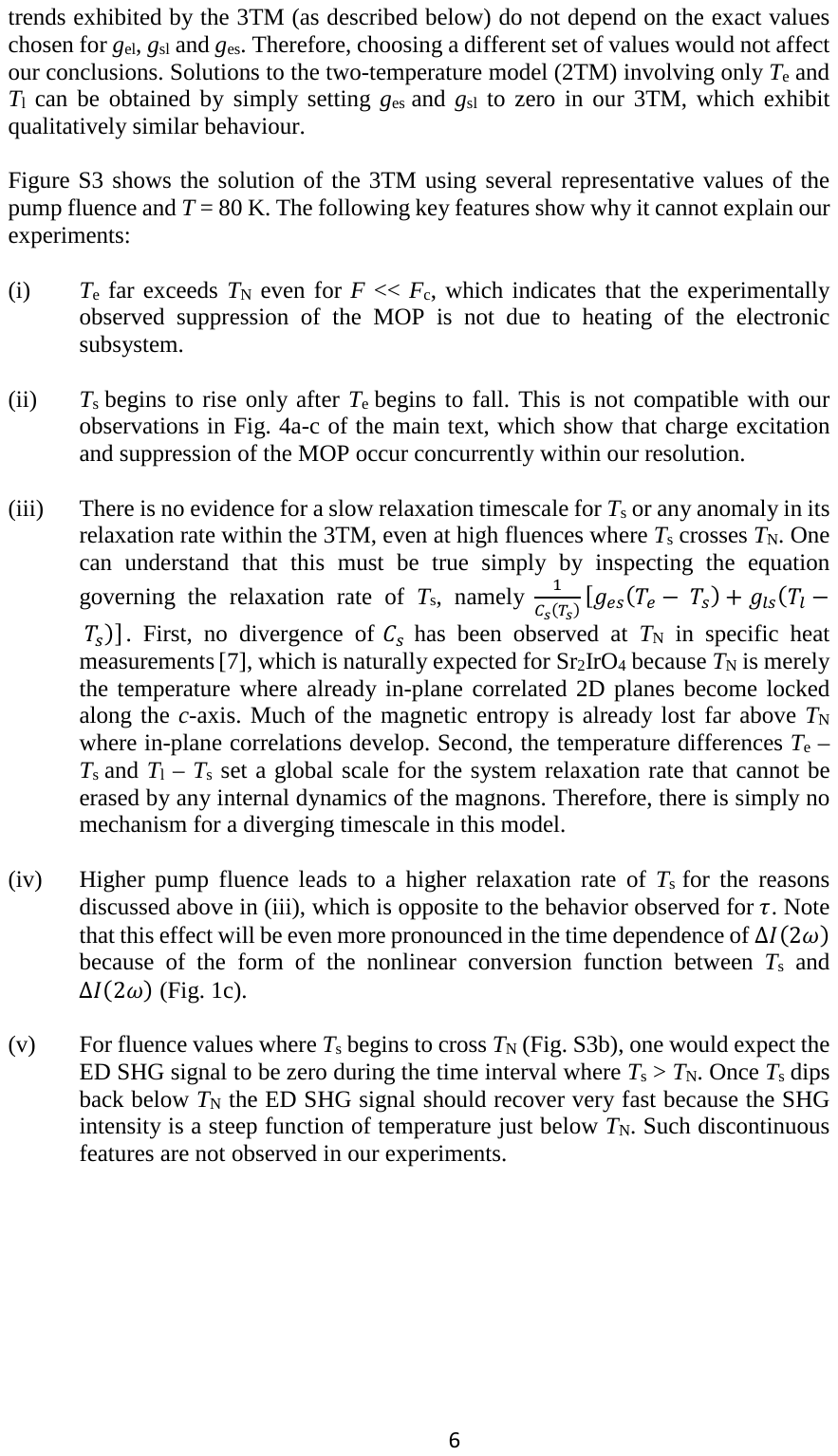}

\end{figure}
\begin{figure}
   \vspace*{-2cm}
   \hspace*{-2cm}
    \centering
    \includegraphics[page=7]{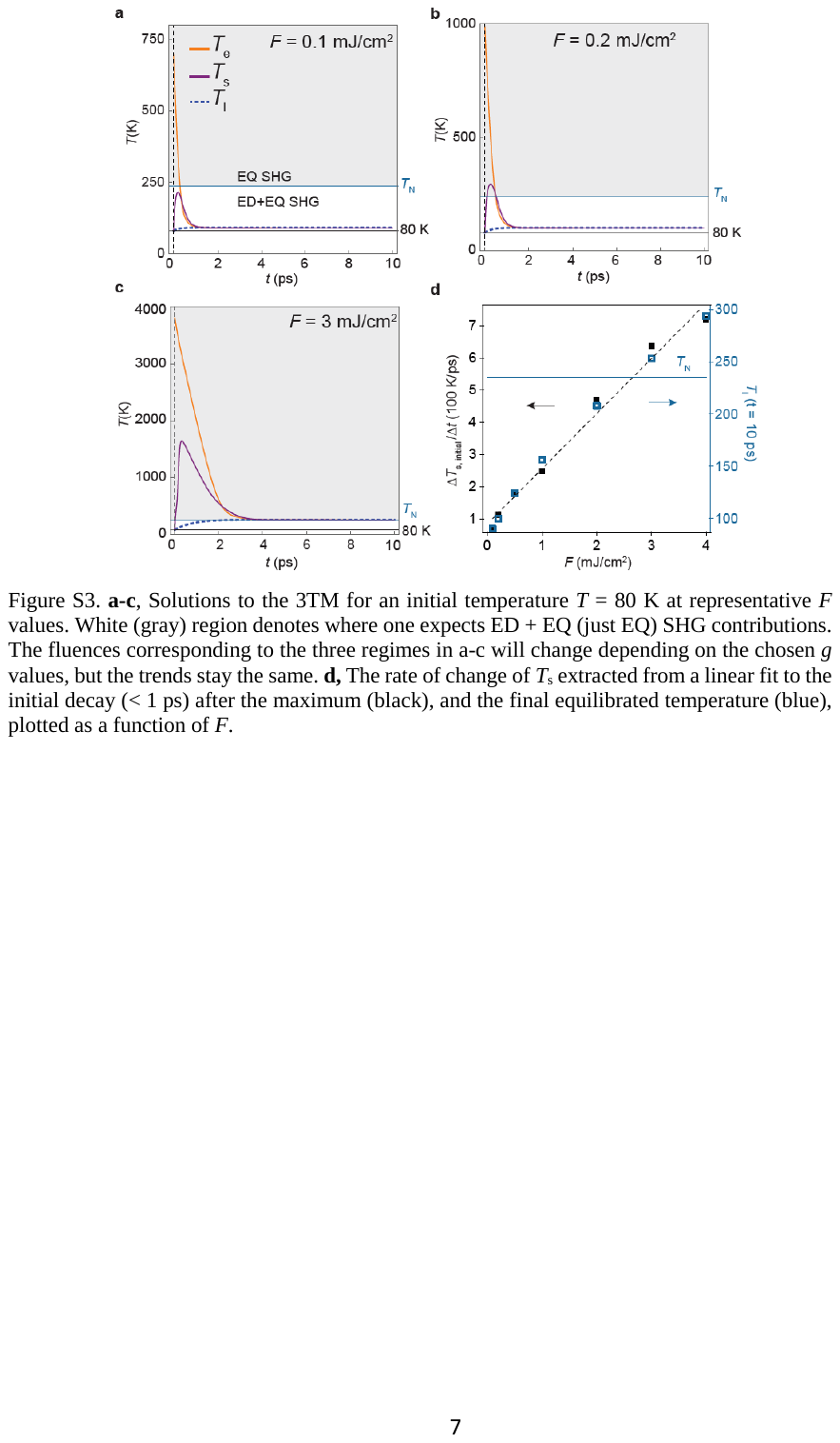}

\end{figure}
\begin{figure}
   \vspace*{-2cm}
   \hspace*{-2cm}
    \centering
    \includegraphics[page=8]{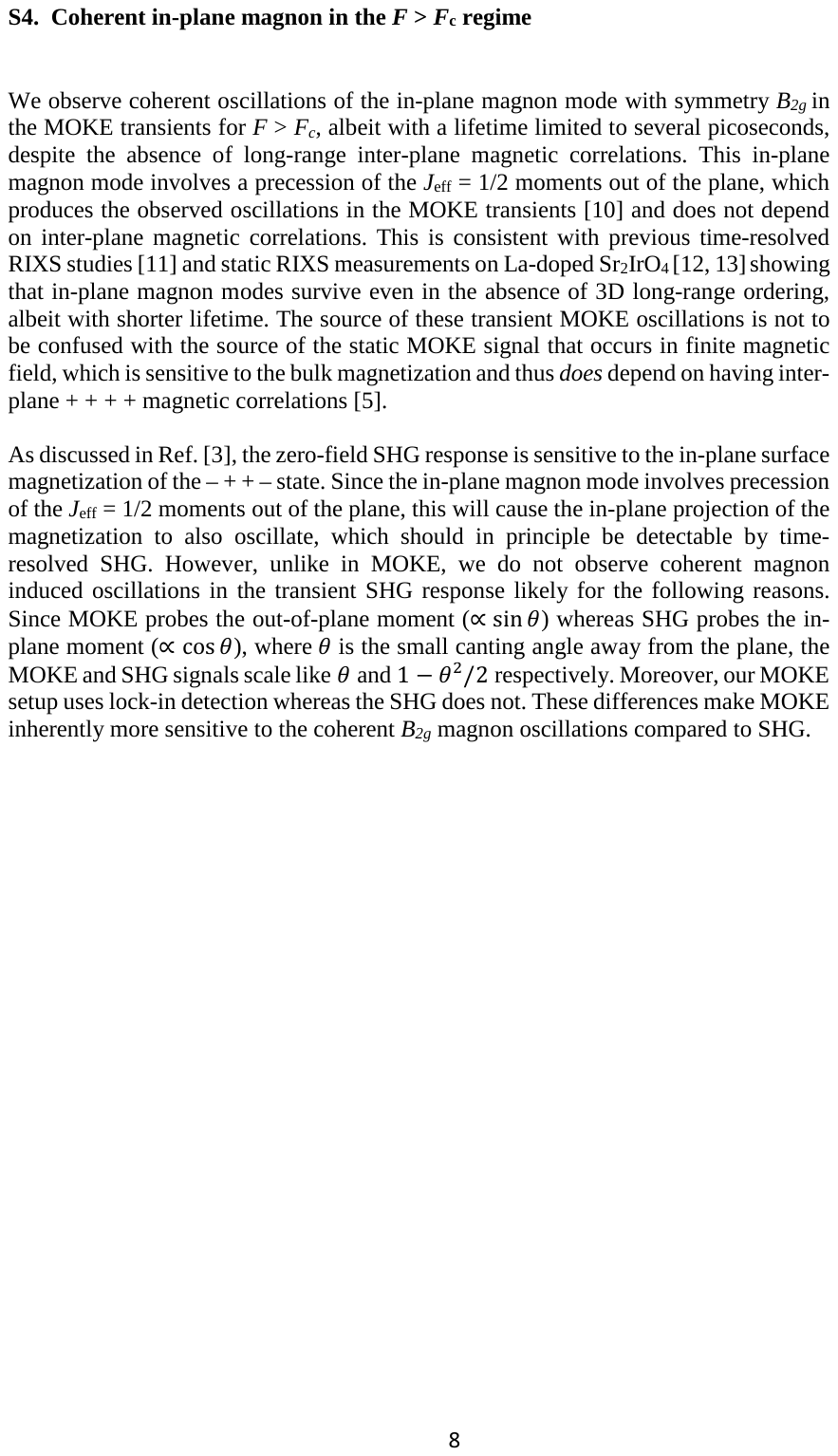}

\end{figure}\begin{figure}
   \vspace*{-2cm}
   \hspace*{-2cm}
    \centering
    \includegraphics[page=9]{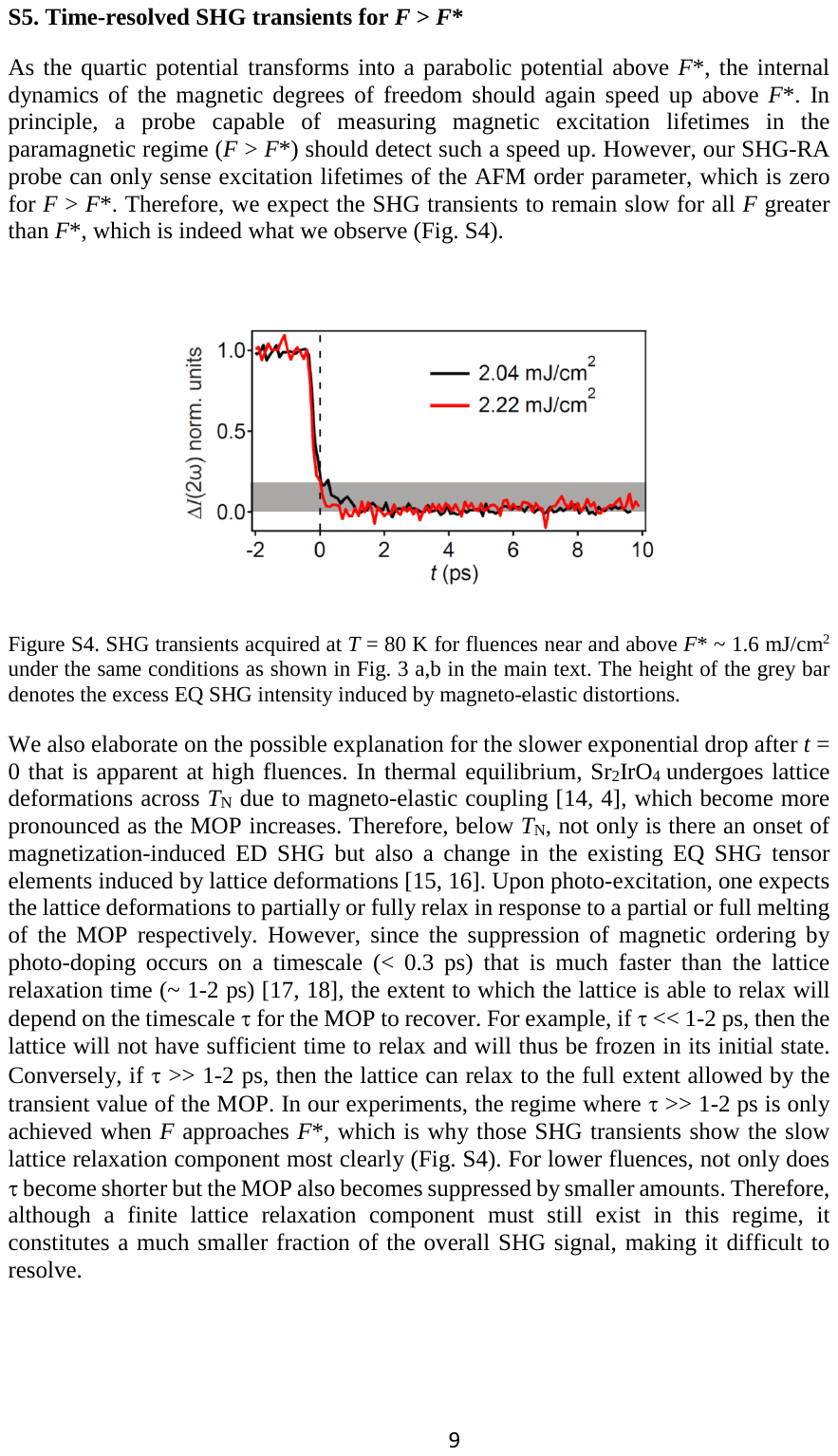}

\end{figure}
\begin{figure}
   \vspace*{-2cm}
   \hspace*{-2cm}
    \centering
    \includegraphics[page=10]{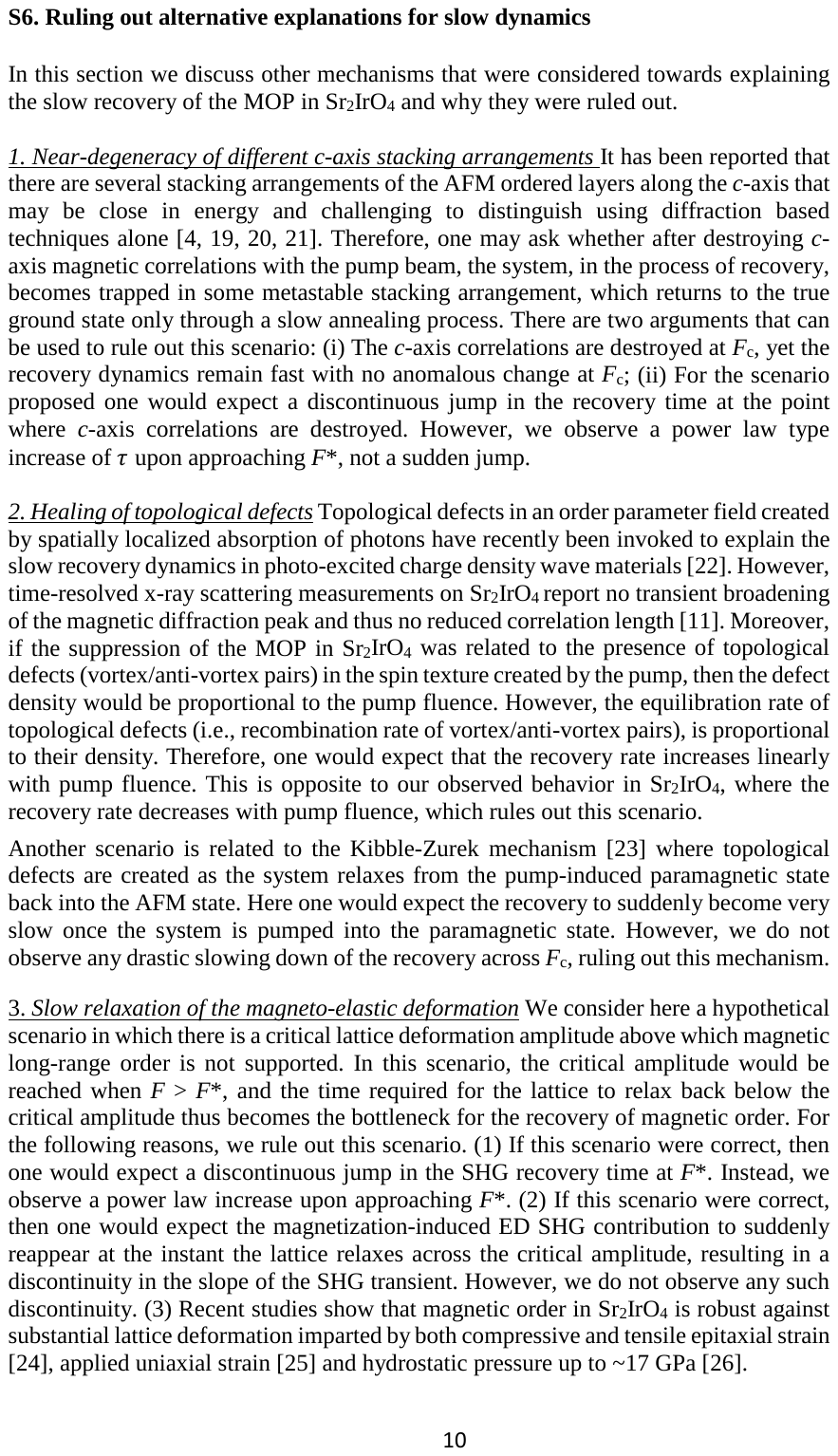}

\end{figure}
\begin{figure}
   \vspace*{-2cm}
   \hspace*{-2cm}
    \centering
    \includegraphics[page=11]{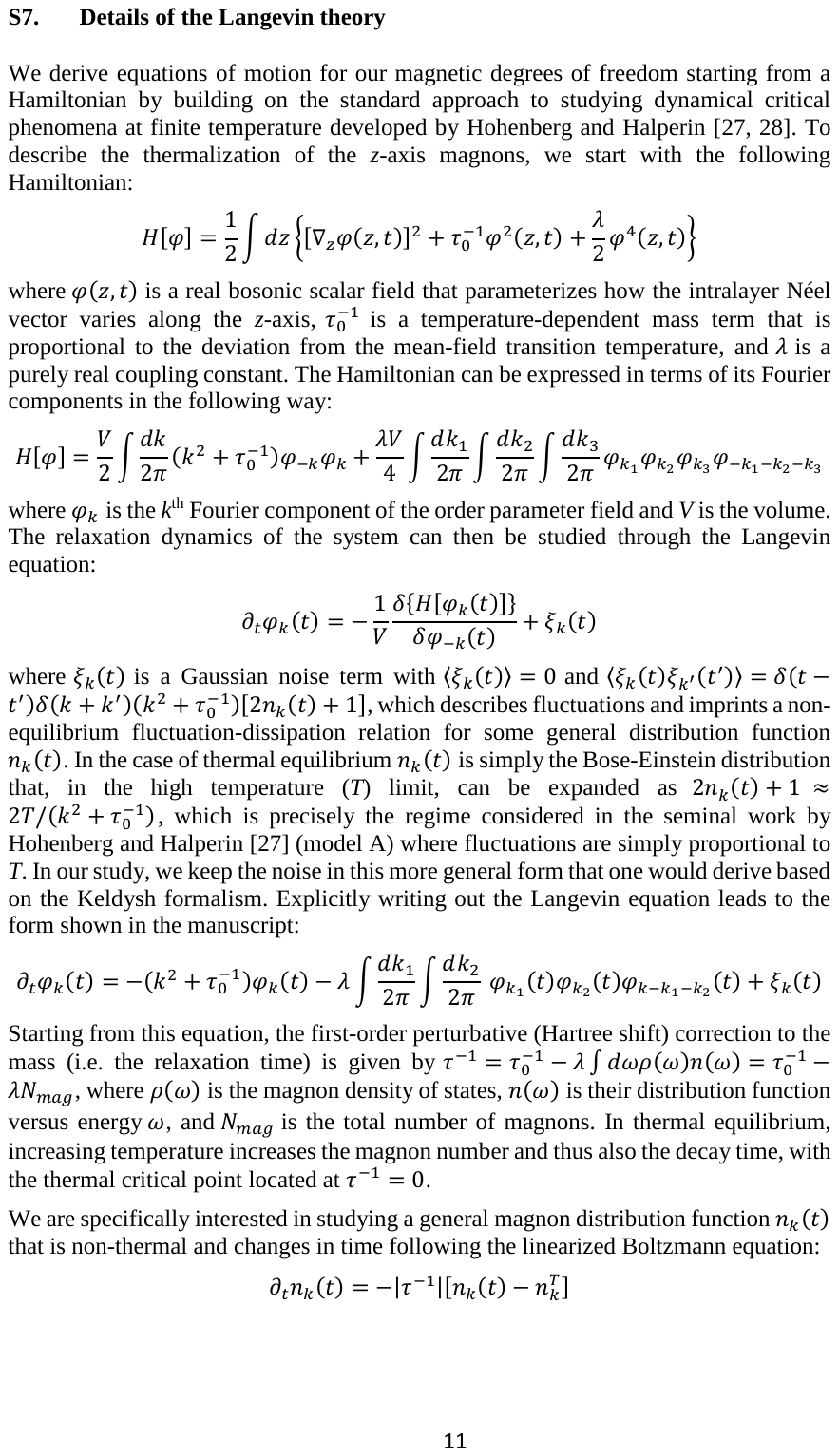}

\end{figure}
\begin{figure}
   \vspace*{-2cm}
   \hspace*{-2cm}
    \centering
    \includegraphics[page=12]{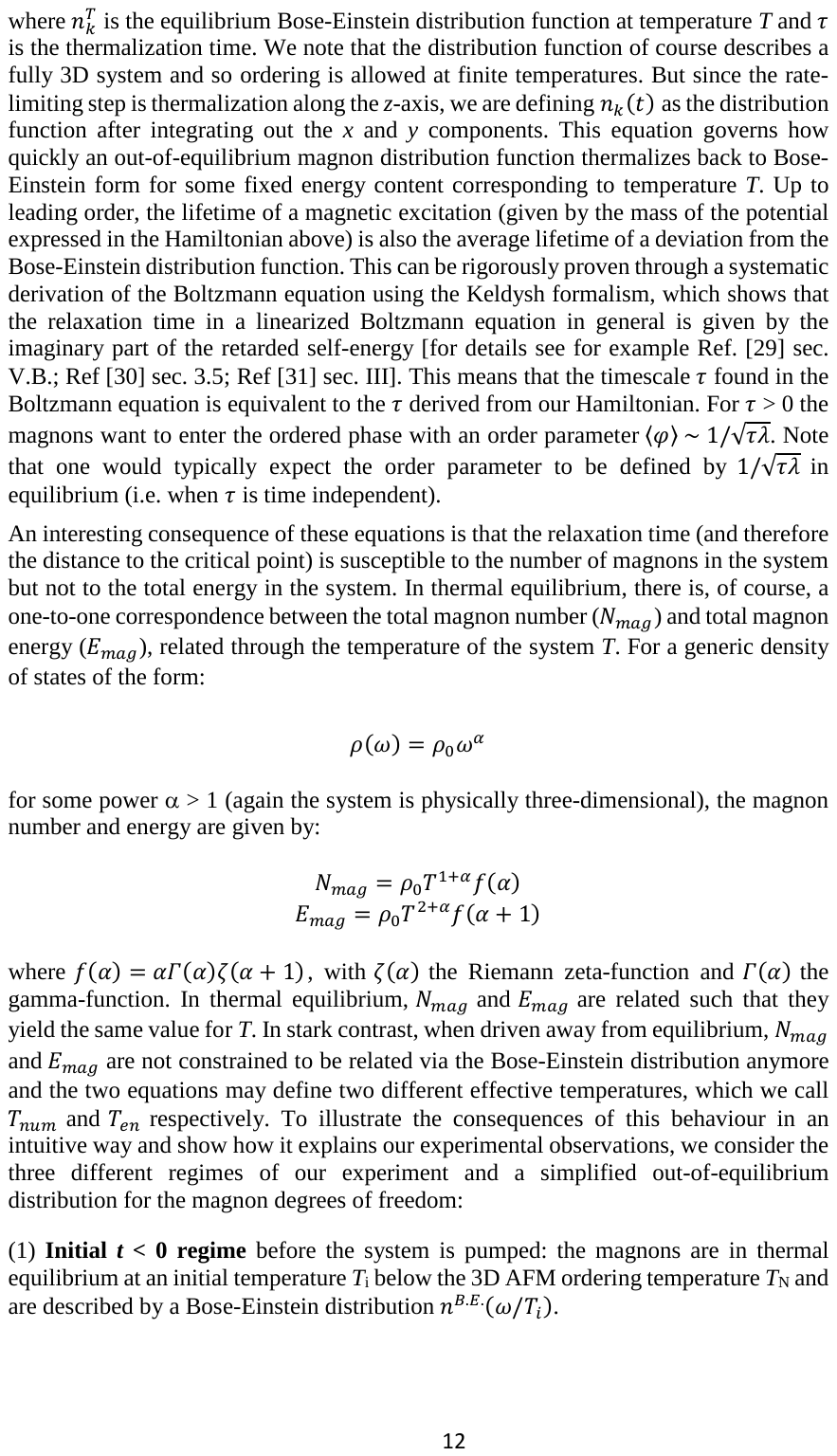}

\end{figure}
\begin{figure}
   \vspace*{-2cm}
   \hspace*{-2cm}
    \centering
    \includegraphics[page=13]{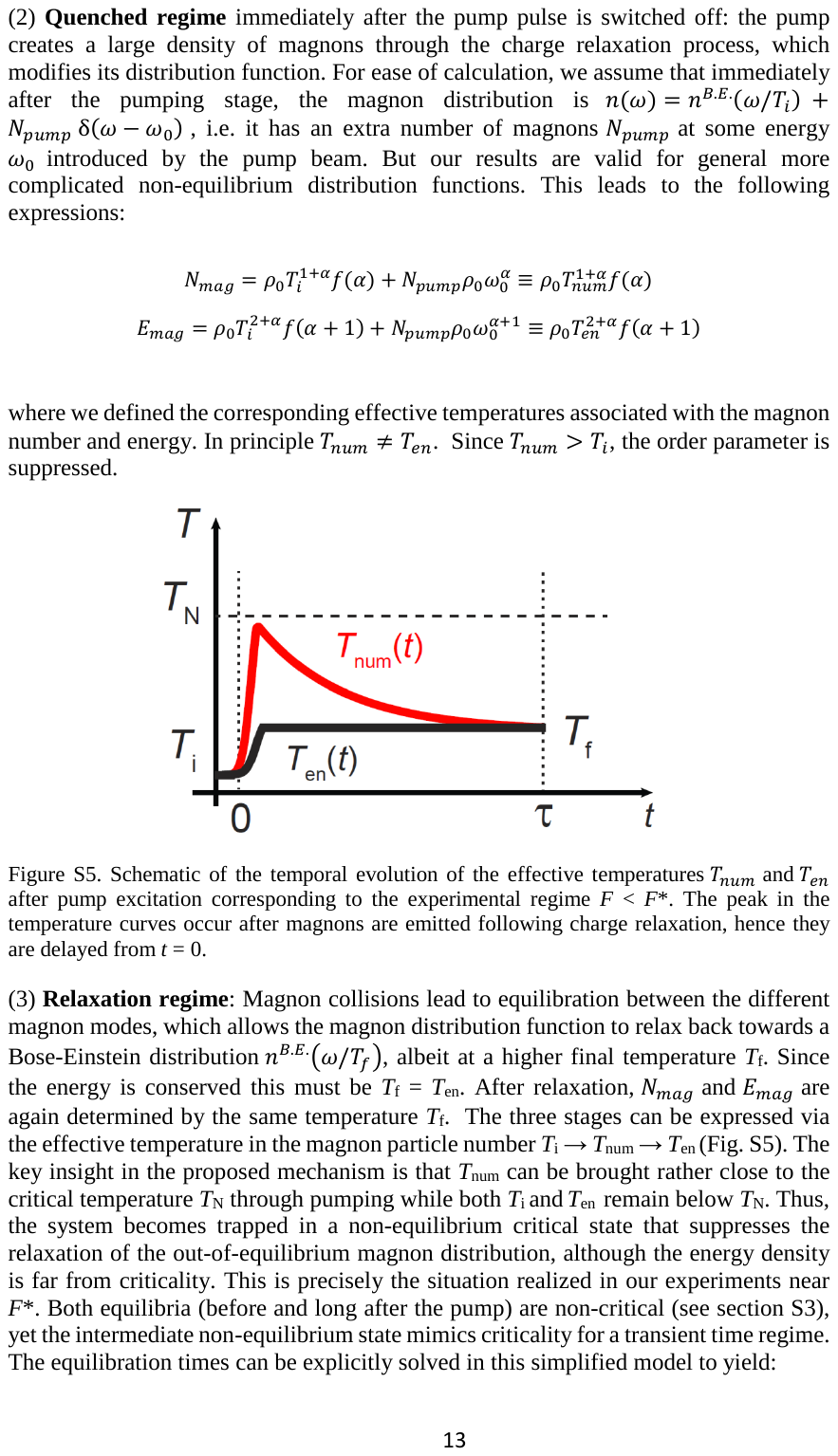}

\end{figure}
\begin{figure}
   \vspace*{-2cm}
   \hspace*{-2cm}
    \centering
    \includegraphics[page=14]{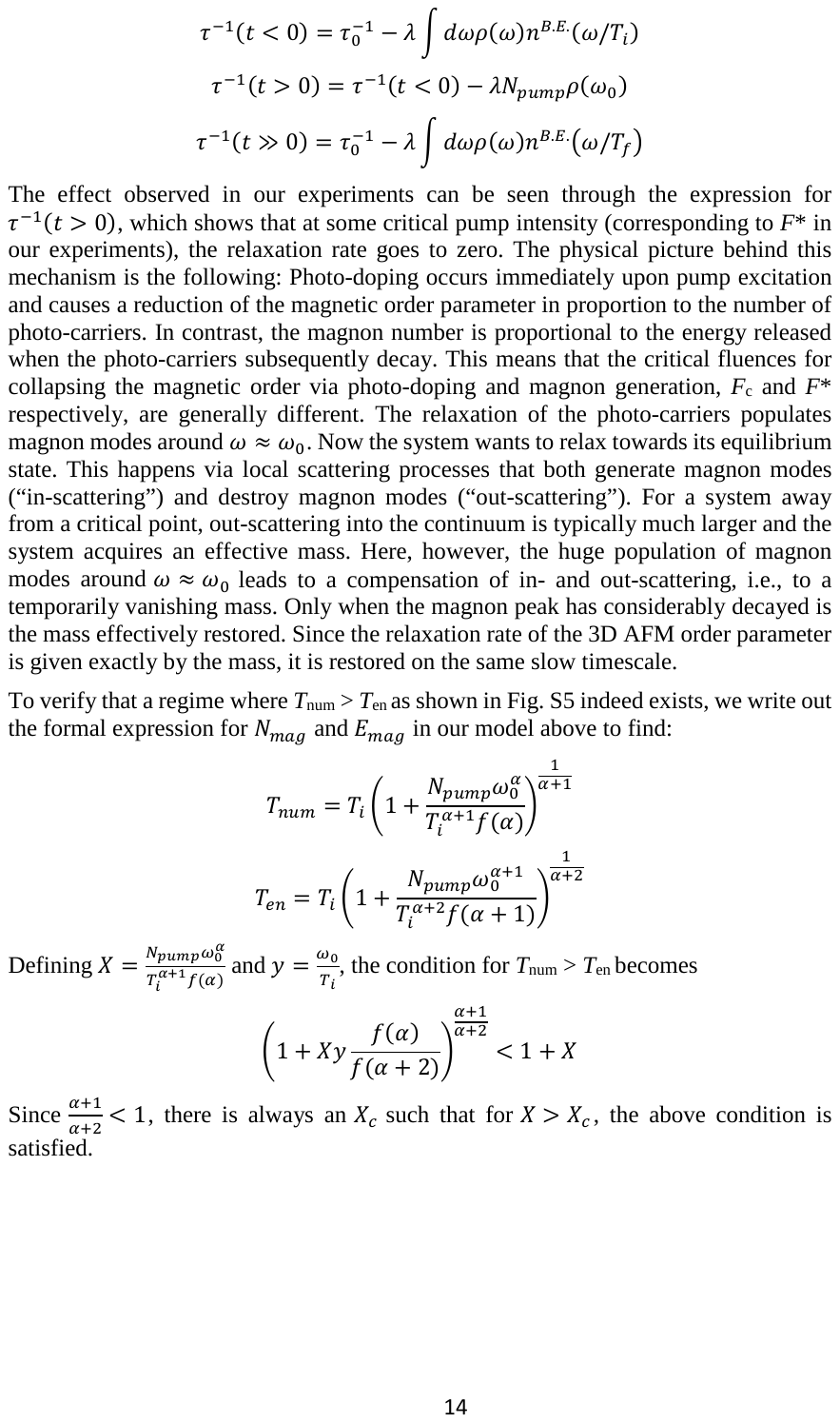}

\end{figure}
\begin{figure}
   \vspace*{-2cm}
   \hspace*{-2cm}
    \centering
    \includegraphics[page=15]{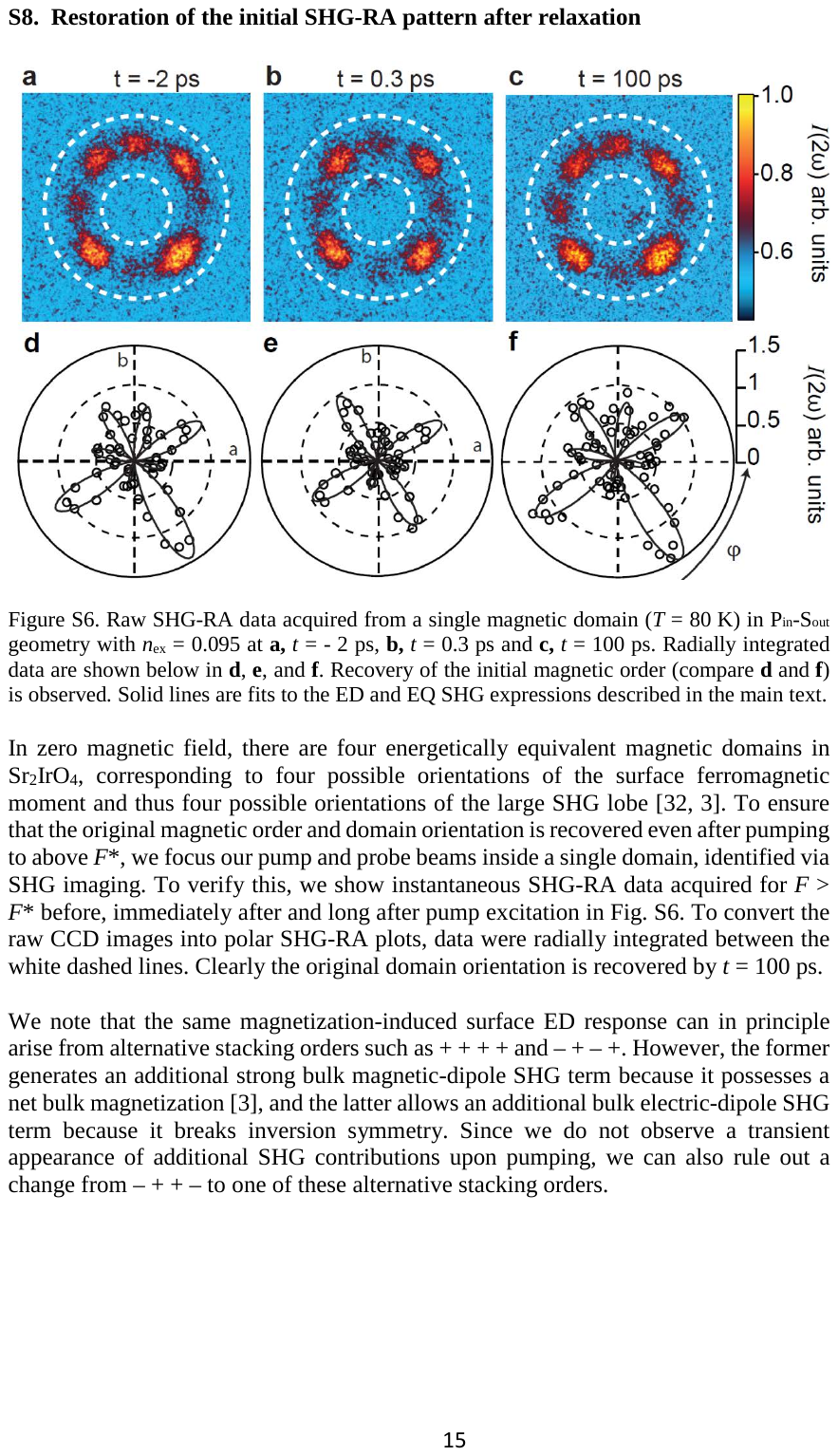}

\end{figure}
\begin{figure}
   \vspace*{-2cm}
   \hspace*{-2cm}
    \centering
    \includegraphics[page=16]{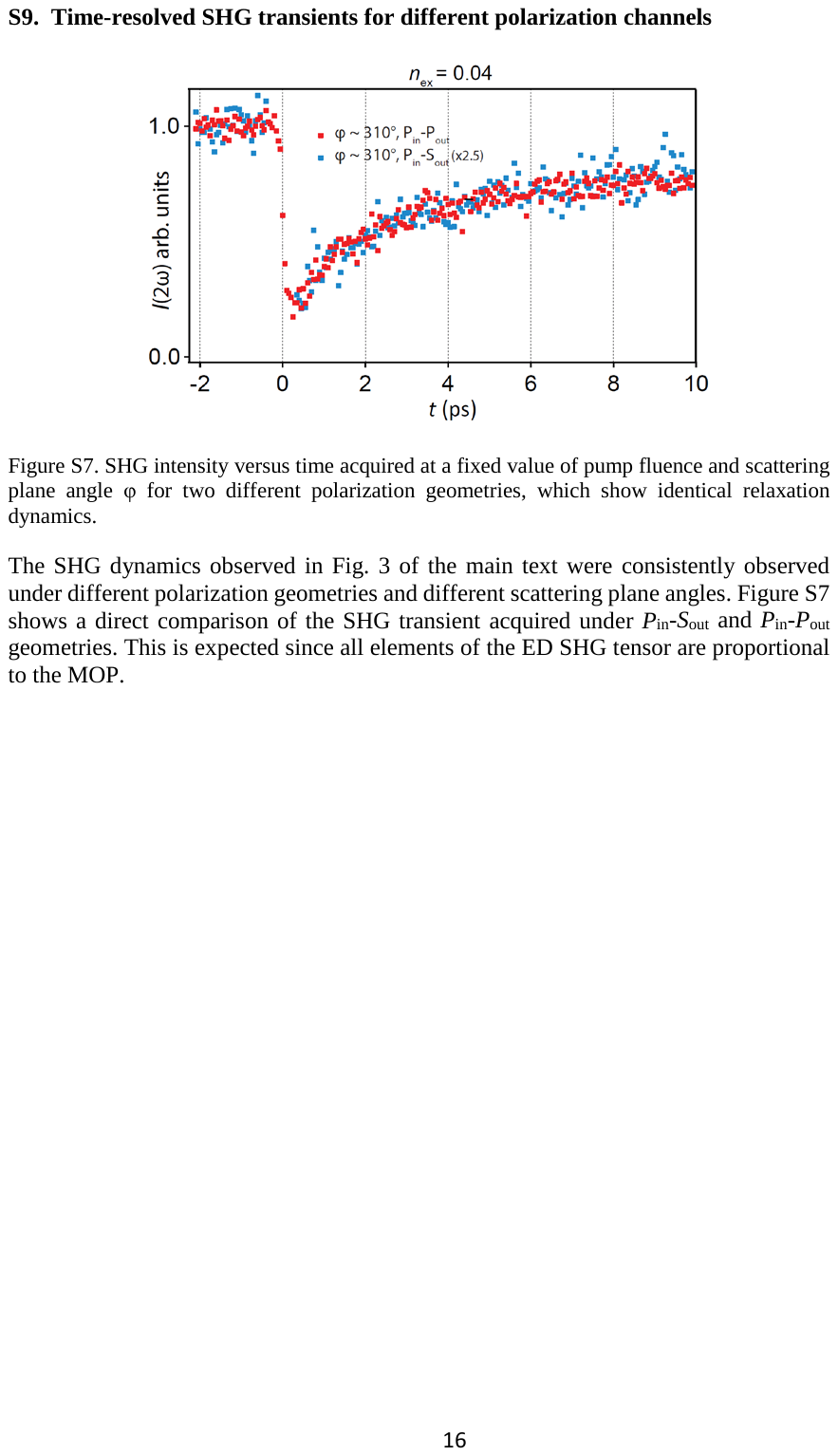}

\end{figure}
\begin{figure}
   \vspace*{-2cm}
   \hspace*{-2cm}
    \centering
    \includegraphics[page=17]{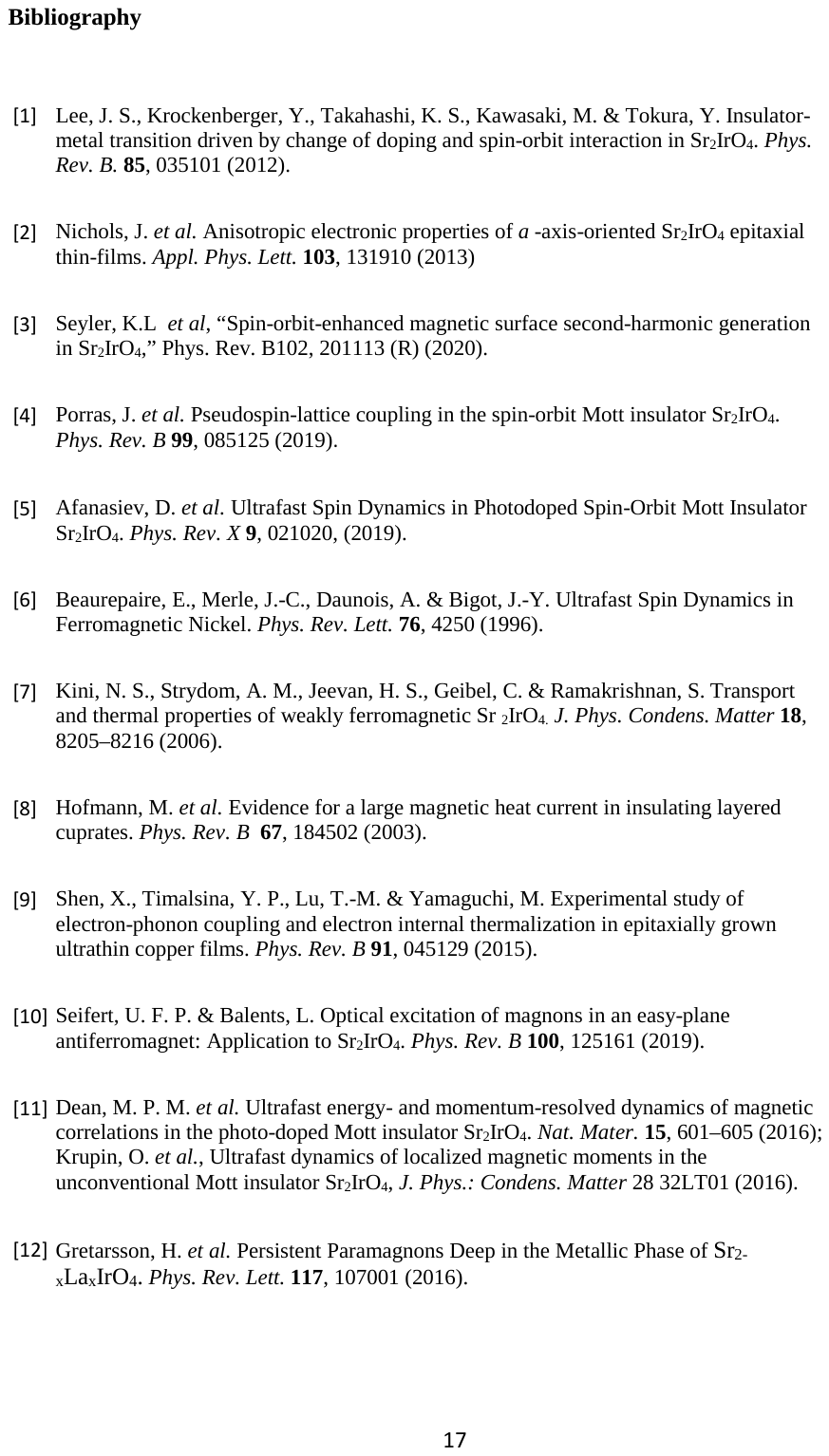}

\end{figure}
\begin{figure}
   \vspace*{-2cm}
   \hspace*{-2cm}
    \centering
    \includegraphics[page=18]{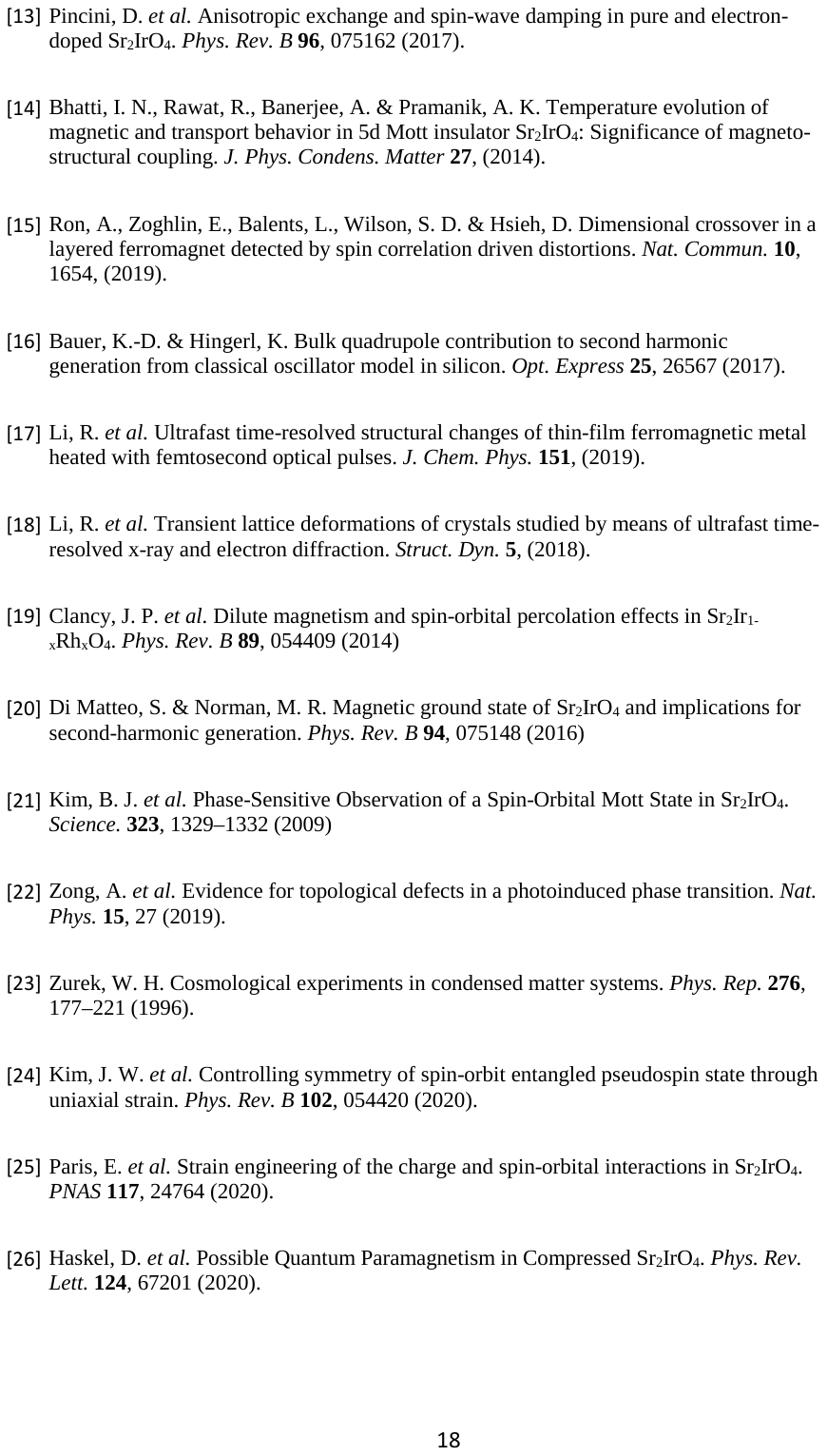}

\end{figure}
\begin{figure}
   \vspace*{-2cm}
   \hspace*{-2cm}
    \centering
    \includegraphics[page=19]{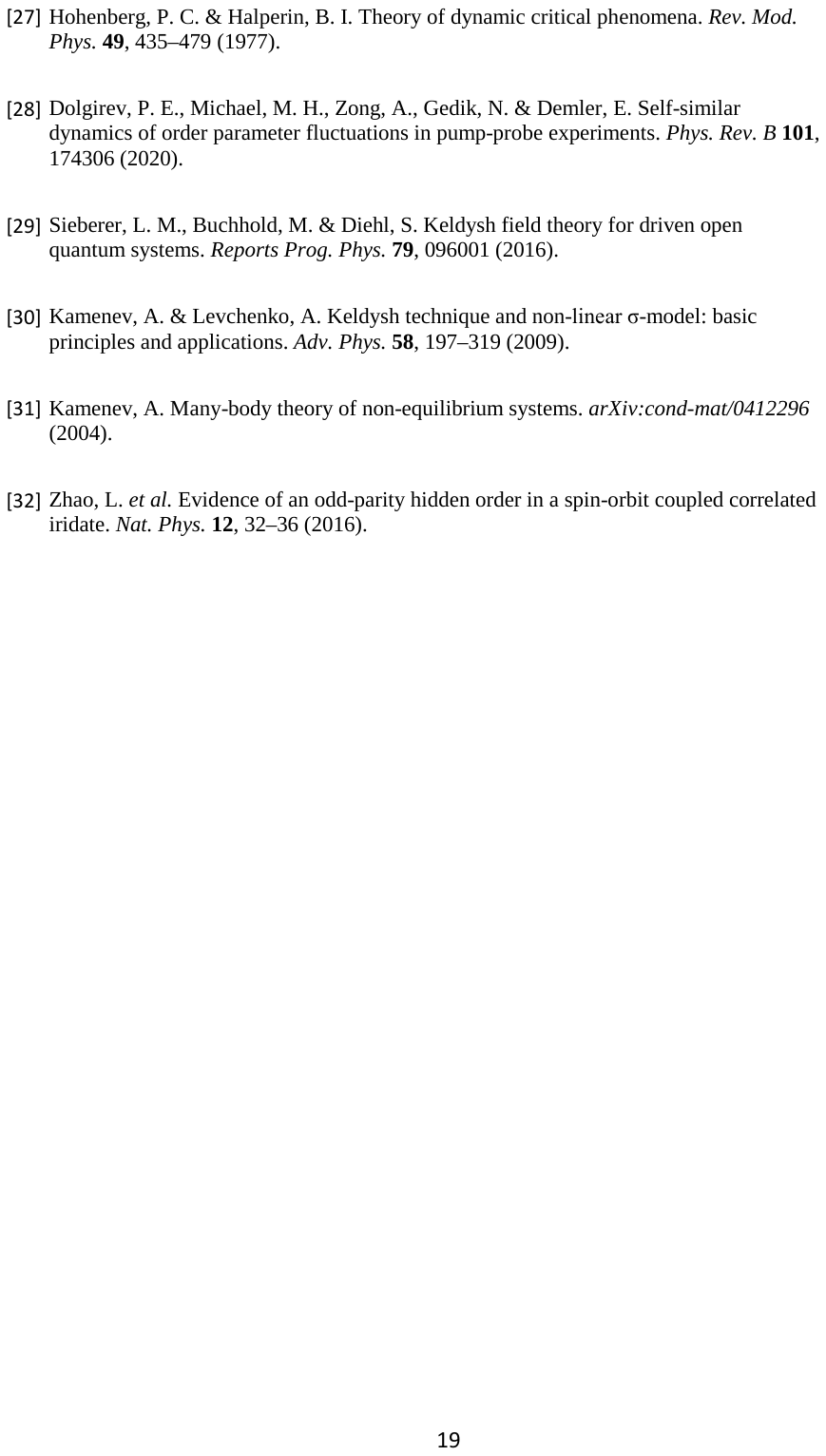}

\end{figure}
\end{document}